\documentclass[journal]{IEEEtran}
\usepackage{cite}
\usepackage{graphicx}
\usepackage{amsmath} 
\usepackage{amssymb} 
\usepackage{subfigure}  
\usepackage{multirow} 
\usepackage{enumerate}
\usepackage{amsthm} 

\begin{document}
\title{Throughput and Delay Analysis of Slotted Aloha with Batch Service}

\author{Huanhuan~Huang,
        Tong~Ye,~\IEEEmembership{Member,~IEEE,}
        and~Tony~T.~Lee,~\IEEEmembership{Fellow,~IEEE}
\thanks{This work was supported in part by the National Natural Science Foundation of
China under Grant 61671286, Grant 61571288, and Grant 61433009.} 
\thanks{H. Huang and T. Ye are with the State Key Laboratory of Advanced Optical Communication Systems and Networks, Shanghai Jiao Tong University, Shanghai 200240, China (email: huanghuanhuan@sjtu.edu.cn; yetong@sjtu.edu.cn).} 
\thanks{T. T. Lee was with the State Key Laboratory of Advanced Optical Communication Systems and Networks, Shanghai Jiao Tong University, Shanghai 200240, China. He is now with the School of Science and Engineering, Chinese University of Hong Kong (Shenzhen), Shenzhen, China (e-mail: tonylee@cuhk.edu.cn)}}

\markboth{}%
{}
\maketitle

\begin{abstract}
In this paper, we study the throughput and delay performances of the slotted Aloha with batch service, which has wide applications in random access networks. Different from the classical slotted Aloha, each node in the slotted Aloha with batch service can transmit up to $M$ packets once it succeeds in channel competition. The throughput is substantially improved because up to $M$ packets jointly undertake the overhead due to contention. In an innovative vacation model developed in this paper, we consider each batch of data transmission as a busy period of each node, and the process between two successive busy periods as a vacation period. We then formulate the number of arrivals during a vacation period in a renewal-type equation, which characterizes the dependency between busy periods and vacation periods. Based on this formulation, we derive the mean waiting time of a packet and the bounded delay region for the slotted Aloha with batch service. Our results indicate the throughput and delay performances are substantially improved with the increase of batch size $M$, and the bounded delay region is enlarged accordingly. As $M$ goes to infinity, we find the saturated throughput can approach $100\%$ of channel capacity, and the system remains stable irrespective of the population size and transmission probability.
\end{abstract}


\section{Introduction}
\IEEEPARstart{S}{lotted} Aloha is a medium access control (MAC) protocol designed for wireless multiple access networks. The slotted Aloha is easy to implement and can provide low-access delay when the traffic load is small \cite{MTC2017}. Due to such advantages, the slotted Aloha or slotted Aloha-like protocols \cite{SA2017ICC,FASA2013TON,p-persistent2012,USN2011VTC,802.15.6.2015,802.15.6standard} have been widely applied in different scenarios, such as Machine-to-Machine (M2M) networks \cite{SA2017ICC, FASA2013TON}, underwater acoustic networks \cite{p-persistent2012,USN2011VTC}, and wireless body area networks (WBANs) \cite{802.15.6.2015,802.15.6standard}.

However, the throughput of the slotted Aloha is quite low when the traffic load is moderate or high, which limits its application in next-generation wireless access networks. In the slotted Aloha, each backlogged node transmits a packet at the beginning of a time slot with a preset probability, called the transmission probability in \cite{2005stability}. If exactly one node attempts to transmit in this slot, this node will transmit the packet successfully; otherwise, multiple packets will collide with each other and no transmission will be successful. No matter what happens in this slot, all the backlogged nodes repeat the procedure in the next slot. Due to frequent collisions, the maximum throughput of the slotted Aloha is only 0.368 \cite{kleinrock1973}.

To enhance the throughput of the slotted Aloha, Ref. \cite{SA2017distributed,Bayesian2013,2001pseudoBayesian,SA2017TON,2005BackoffTON} proposed to dynamically adjust the transmission probability of the node according to the network state. In \cite{SA2017distributed,Bayesian2013,2001pseudoBayesian}, the node lowers the transmission probability of each node when the number of backlogged nodes is large. In \cite{SA2017TON,2005BackoffTON}, the node reduces the transmission probability exponentially with the number of collisions it has involved. However, such dynamic adjustment of the transmission probability can only offer marginal improvement of the throughput. For example, the maximum throughput of the scheme in \cite{SA2017TON} is only 0.43. This is attributed to the fact that each packet in these schemes still involves channel competition before successful transmission.

To avoid excessive collisions, Ref. \cite{M2M2018Dai,jiang2018mIoT,T-Lohi} introduce the concept of batch service to the slotted Aloha. The general idea of batch service is that the node can send a batch of packets once it makes a successful attempt, such that the overhead incurred by channel competition is shared by multiple packets. An example is a protocol called T-Lohi \cite{T-Lohi}, proposed for underwater acoustic communications. In this protocol, a node can transmit multiple packets once it succeeds in the competition process. This way, not every packet needs to experience the competition process before successful transmission so that many collisions can be avoided. Cellular-based IoTs are another examples \cite{M2M2018Dai,jiang2018mIoT}. In the IoT network, each device sends requests to the base station in a random manner. Once the base station receives the request successfully, the related device can send out the packets. Recently, Ref. \cite{Dai2019random} demonstrated that the throughput can be substantially improved in this way. It is also reported in \cite{T-Lohi} that the maximal throughput of the slotted Aloha with batch service is much larger than that of the classical slotted Aloha.

Despite the elegance of the slotted Aloha with batch service, its performance is not fully understood. For example, how are the network throughput and the mean delay of packets influenced by the system parameters, such as transmission probability and batch size? What is the maximal throughput? What is the condition of bounded mean delay? To answer these questions, an analytical model is indispensable.

\subsection{Previous Works}
The study of classical slotted Aloha originated in the 1970s \cite{kleinrock1973,abramson1973}. These works mainly focused on the throughput analysis of saturated networks where each node always had packets to send. Assuming the number of attempts in each slot as a Poisson random variable with a mean value of $G$, Ref. \cite{kleinrock1973,abramson1973} shows that the saturated throughput is $Ge^{-G}$, which reaches the maximum value $e^{-1}$ when $G=1$. However, the network is unsaturated in practice. In this case, the mean delay of the packet is also an important criterion to measure network performance. Though the model in \cite{kleinrock1973,abramson1973} can characterize the contention behavior among nodes well, it cannot describe the queueing behavior of packets when the network is unsaturated.

Ref. \cite{SA2017TON,1999stability,1988stability,1979ergodicity} then studied the slotted Aloha in unsaturated networks. Ref. \cite{1999stability,1988stability,1979ergodicity} delineated an $n$-node network as an $n$-dimensional Markov chain, of which the state space is too large to be solved when $n>2$. Ref. \cite{SA2017TON} investigated the slotted Aloha with an exponential backoff mechanism and modeled the system as an infinite-dimensional Markov chain, in which the system state was defined by the number of nodes in each backoff stage. Again, this model was too complex to be solved, and only some special cases were tractable. The difficulty of the models in \cite{SA2017TON,1999stability,1988stability,1979ergodicity} lies in that these models tried to take into consideration the dependencies among the queues of all the nodes.

Ref. \cite{dai2012stability} focused on networks with a large number of nodes, in which the dependency among the queues of different nodes becomes so weak that each queue can be analyzed independently and separately \cite{Dai2013CSMA,sun2017fairness}. Ref. \cite{dai2012stability} modeled each node as a Geo/G/1 queue, where the service time of each packet was defined as the duration from the epoch when it became the head-of-line (HOL) packet to the time when it was successfully transmitted. The key step is to use a Markov chain to delineate the attempt process of the HOL packet, from which Ref. \cite{dai2012stability} derived the service time distribution and finally obtained the mean delay. Based on the analytical results, Ref. \cite{dai2012stability} discussed the stable region of the classical slotted Aloha.

However, the model in \cite{dai2012stability} cannot be used to analyze the slotted Aloha with batch service. Unlike that in the classical slotted Aloha, the node in the slotted Aloha with batch service can send multiple packets after each successful attempt. This implies that different packets have different behaviors. In particular, each HOL packet at the time when the node takes over the channel has to experience the competition process before successful transmission, while those packets waiting in the buffer at that time can be transmitted directly without any attempt. Therefore, the competition process cannot be regarded as a part of the service of a packet, when we analyze the slotted Aloha with batch service.

\subsection{Our Contributions}
In this paper, we develop a generalized model to analyze the slotted Aloha with batch service, where each node can transmit up to $M$ packets once it succeeds in the channel competition. Our goal is to study the delay performance and the stability condition of this protocol and show how batch service discipline can improve performance.

Different from \cite{dai2012stability}, we model each node as a queue with vacation, where we consider each batch of data transmission as a busy period, while the attempt process between two successive busy periods is considered a vacation period. We show that the vacation period is controlled by the packet arrival process and the competition process, in which the successful probability of each node is related to the mean number of attempts per time slot, called the attempt rate in this paper. We thus start our analysis with the derivation of the distribution of the number of arrivals during the vacation period and the attempt rate. We derive such a distribution using renewal equations, which are built according to the feature of the channel competition process, and the attempt rate using a Lindley's equation, which is established based on the batch-service principle. We then obtain the mean waiting time and the bounded delay region in terms of the transmission probability $r$.

Our results clearly indicate that, with the increase of batch size $M$, the throughput and delay performance are improved, and the bounded delay region is enlarged accordingly. Especially, when $M$ is infinity, the bounded delay region is the whole region of $r$, i.e., (0, 1) for any node population and any aggregate input traffic rate smaller than 1. This indicates the batch-service slotted Aloha with $M=\infty$ is quite robust with respect to $r$.
In summary, the contributions of this paper are as follows:
\begin{enumerate}[{(1)}]
\item
We build a generalized vacation model to analyze the slotted Aloha with batch service, in which the vacation period is governed by the arrival process and the channel competition process.
\item
We obtain the closed-form expression for the mean waiting time of the slotted Aloha with batch service, based on which we theoretically demonstrate that the batch-service discipline can substantially improve the delay performance of the slotted Aloha.
\end{enumerate}

The rest of this paper is organized as follows. Section \uppercase\expandafter{\romannumeral2} presents the working process of slotted Aloha with batch service, and analyzes the network throughput. Section \uppercase\expandafter{\romannumeral3} devises a vacation model to delineate the queueing behavior of each node and analyzes the distribution of the number of arrivals during a vacation period. Section \uppercase\expandafter{\romannumeral4} derives the expression of the attempt rate. Section \uppercase\expandafter{\romannumeral5} discusses the mean waiting time of packets and the bounded delay region. Section \uppercase\expandafter{\romannumeral6} concludes this paper.

\section{Overview and Preliminary Throughput Analysis}
Different from the classical slotted Aloha, in which a node releases the channel immediately after a successful transmission of a HOL packet, a node in the slotted Aloha with batch service continuously sends at most $M$ packets that arrived before the successful transmission of the first HOL packet. Since predominate possible collisions can be avoided, except for the first HOL packet, a significant amount of overhead incurred by the channel competition could be eliminated.

The slotted Aloha with batch service studied in this paper is described as follows. If the channel is free in a slot, each backlogged node attempts to send a packet with transmission probability, denoted by $r$. Each node maintains a virtual gate in the buffer. At the beginning of the slot in which the node makes an attempt, the node puts the first $K$ packets inside the gate, where $K$ is equal to the smaller of the queue length and a preset integer $M$, called the batch size in this paper. If a node succeeds, it will take over the channel and send out $K-1$ packets inside the gate. To achieve this function, this node appends one additional bit, called a \emph{reservation bit}, in each packet to inform the access point (AP) whether it has a transmission in the next slot. If this node will send a packet in the next slot, the reservation bit embedded in the current packet is 1; otherwise, it is 0. Once the AP receives the packet, it will broadcast an acknowledgment (ACK) embedded with a reservation bit to all nodes. The ACK not only confirms the successful transmission of the packet, but also informs all the nodes of the channel state in the next slot. If the bit is 1, the channel will still be in use and other nodes will keep silent in the next slot; otherwise, they may attempt to access the channel.

\begin{figure}[!t]
\centering
\includegraphics[scale=0.35]{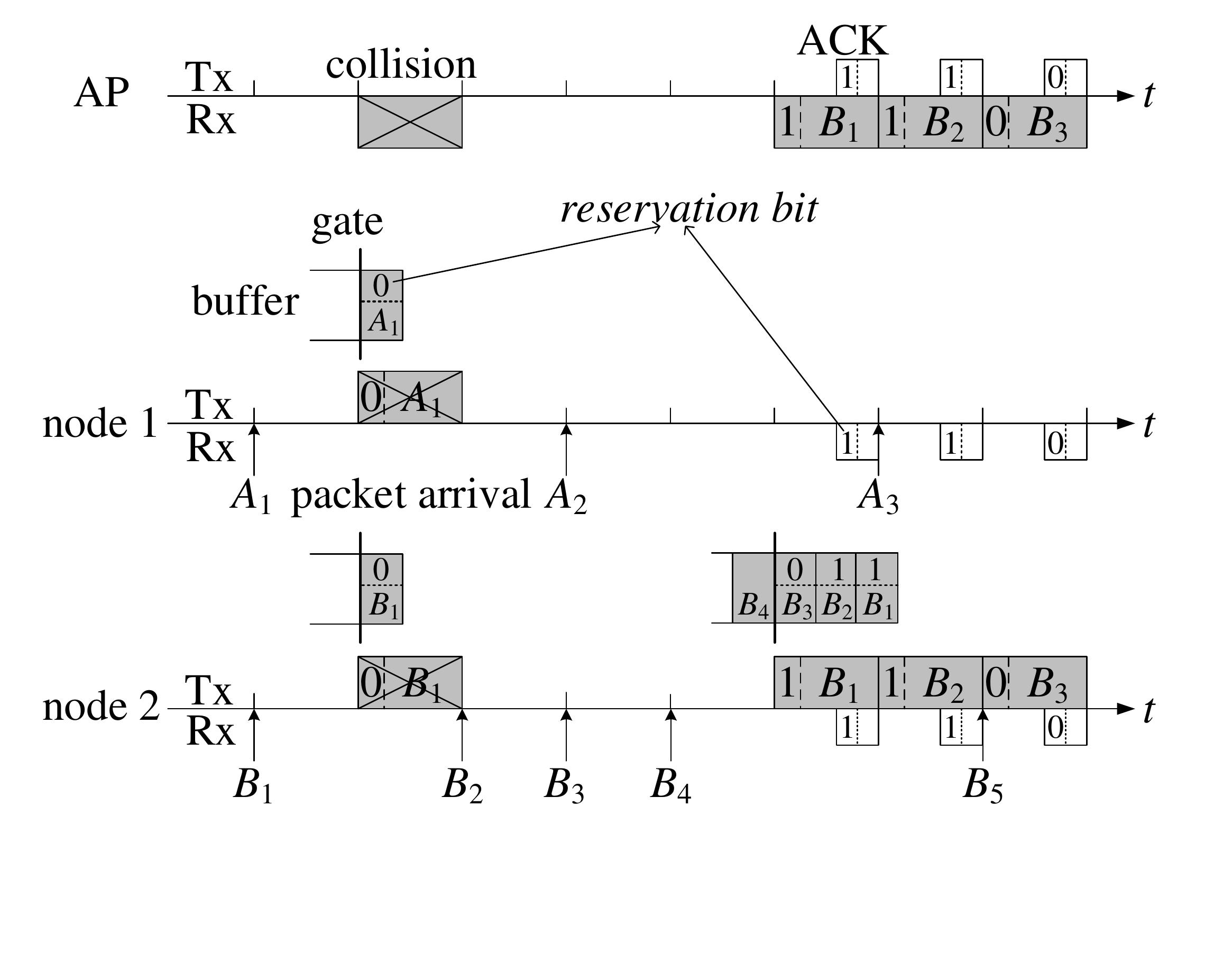}
\caption{Process of slotted Aloha with batch service discipline where $n=2$ and $M=3$.}
\label{fig1}
\end{figure}

Fig.~\ref{fig1} illustrates an example where there are two nodes and $M=3$. At the beginning of the first slot, packets $A_1$ and $B_1$ arrive at node 1 and node 2, respectively, such that the queue length of these two nodes is 1. In the second slot, both nodes make attempts. Thus, at the beginning of the second slot, each node puts one packet into the virtual gate and sets the reservation bit of the packet to 0. Of course, a collision happens and packet transmissions fail in the second slot. In the subsequent three slots, packets $B_2$, $B_3$, and $B_4$ arrive at node 2. In the sixth slot, node 2 makes an attempt and puts the first $K=min\{3,4\}=3$ packets into the virtual gate. Node 2 also sets the reservation bits of $B_1$ and $B_2$ to 1 and that of $B_3$ to 0. Hence, after the successful transmission of $B_3$, the channel is released and node 1 and node 2 can compete for the channel again.

To analyze the slotted Aloha with batch service, we adopt the following assumptions throughout this paper:
\begin{itemize}
\item[{A1.}]
The channel is error free;
\item[{A2.}]
The number of nodes, denoted by $n$, is sufficiently large;
\item[{A3.}]
All of $n$ nodes are statistically identical;
\item[{A4.}]
A node has a packet arrival at the beginning of a slot with probability $\lambda$, and thus the aggregate packet arrival rate of the whole network is $\hat{\lambda}=n\lambda$ packets/slot;
\item[{A5.}]
Each node transmits the packets in a first-in-first-out manner;
\item[{A6.}]
The transmission time of each packet is one slot.
\end{itemize}
According to A2, when the channel is free, the number of attempts in a slot is approximately a Poisson random variable \cite{dai2012stability,wong2011CSMA,Dai2013CSMA}, of which the mean is called the attempt rate and denoted by $G$.

\subsection{Saturated Throughput}
\begin{figure}[!t]
\centering
\includegraphics[scale=0.28]{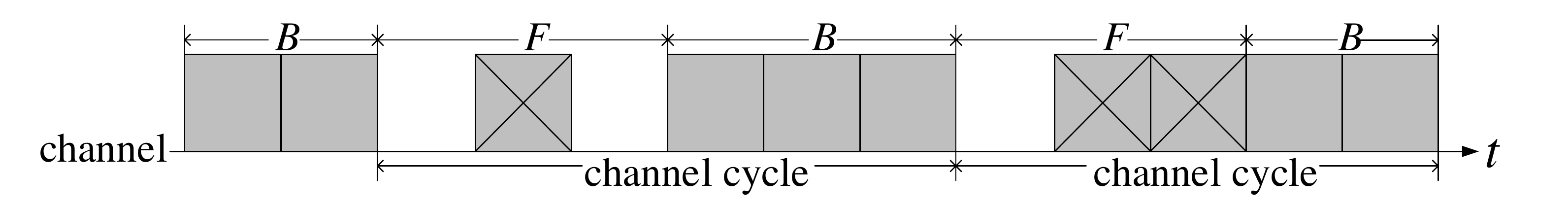}
\caption{The channel cycle in the time axis.}
\label{fig2}
\end{figure}

As Fig.~\ref{fig2} illustrates, the channel works in a cyclic manner. Each channel cycle consists of a competition period, denoted as $F$, followed by a busy period, denoted as $B$. In each slot of the competition period, the channel is free, and all the backlogged nodes will make an attempt with probability $r$. According to assumption A2, the number of attempts in a slot is approximately a Poisson random variable with attempt rate $G$. It follows that the probability that only one node makes an attempt (i.e., succeeds) in a slot is $Ge^{-G}$. Once a node succeeds in a slot, the channel enters the busy period in this slot immediately. Therefore, the mean competition period is
\begin{equation}
\overline{F}{\triangleq}E[F]=1/(Ge^{-G})-1.
\label{mean-competition}
\end{equation}
In the busy period, the channel is captured by a node, and the node can transmit $\overline{B}{\triangleq}E[B]$ packets on average. Therefore, the network throughput, defined as the fraction of the time that the channel spends in successful transmissions, is given by:
\begin{equation}
{\hat{\lambda}}_{out}=\frac{\overline{B}}{\overline{F}+\overline{B}}=\frac{1}{\frac{1/(Ge^{-G})-1}{\overline{B}}+1},
\label{throughput}
\end{equation}
where $\frac{1/(Ge^{-G})-1}{\overline{B}}=\frac{\overline{F}}{\overline{B}}$ is the mean competition period the channel needs to transmit one packet successfully, i.e., the amortized competition overhead from the viewpoint of channel. If the network is saturated, meaning that all the nodes always have packets to transmit, $\overline{B}=M$ and the attempt rate $G=nr$. In this case, the saturated throughput is given by
\begin{equation}
{\hat{\lambda}}_{sat}=\frac{1}{\frac{1/(nre^{-nr})-1}{M}+1},
\label{sat-throughput}
\end{equation}
where $\frac{1/(nre^{-nr})-1}{M}$ is the amortized competition overhead in the saturated network. It is clear that $\hat{\lambda}_{sat}$ is the maximum capacity that the channel can offer. Setting $M=1$ in (\ref{sat-throughput}), we immediately obtain the saturated throughput of the classical slotted Aloha $nre^{-nr}=Ge^{-G}$, which coincides with the analytical result of previous work \cite{kleinrock1973}.

As Section \uppercase\expandafter{\romannumeral1} mentions, the advantage of the slotted Aloha with batch service is that the competition overhead can be amortized by all the packets transmitted in the same batch such that the throughput can be substantially enhanced. Such advantage can be easily observed from (\ref{sat-throughput}). With the growth of the batch size $M$, the amortized competition overhead per packet $\frac{1/(nre^{-nr})-1}{M}$ decreases, which in turn improves the saturated throughput. Especially, if $M$ approaches infinity, this overhead reduces to 0 such that the channel throughput can reach $100\%$ for any $r\in(0,1)$, any $\hat{\lambda}\in(0,1)$, and any node population $n$.
\begin{figure*}[!t]
\centering
\subfigure[$n=30$]{
\label{fig3-a}
\includegraphics[scale=0.37]{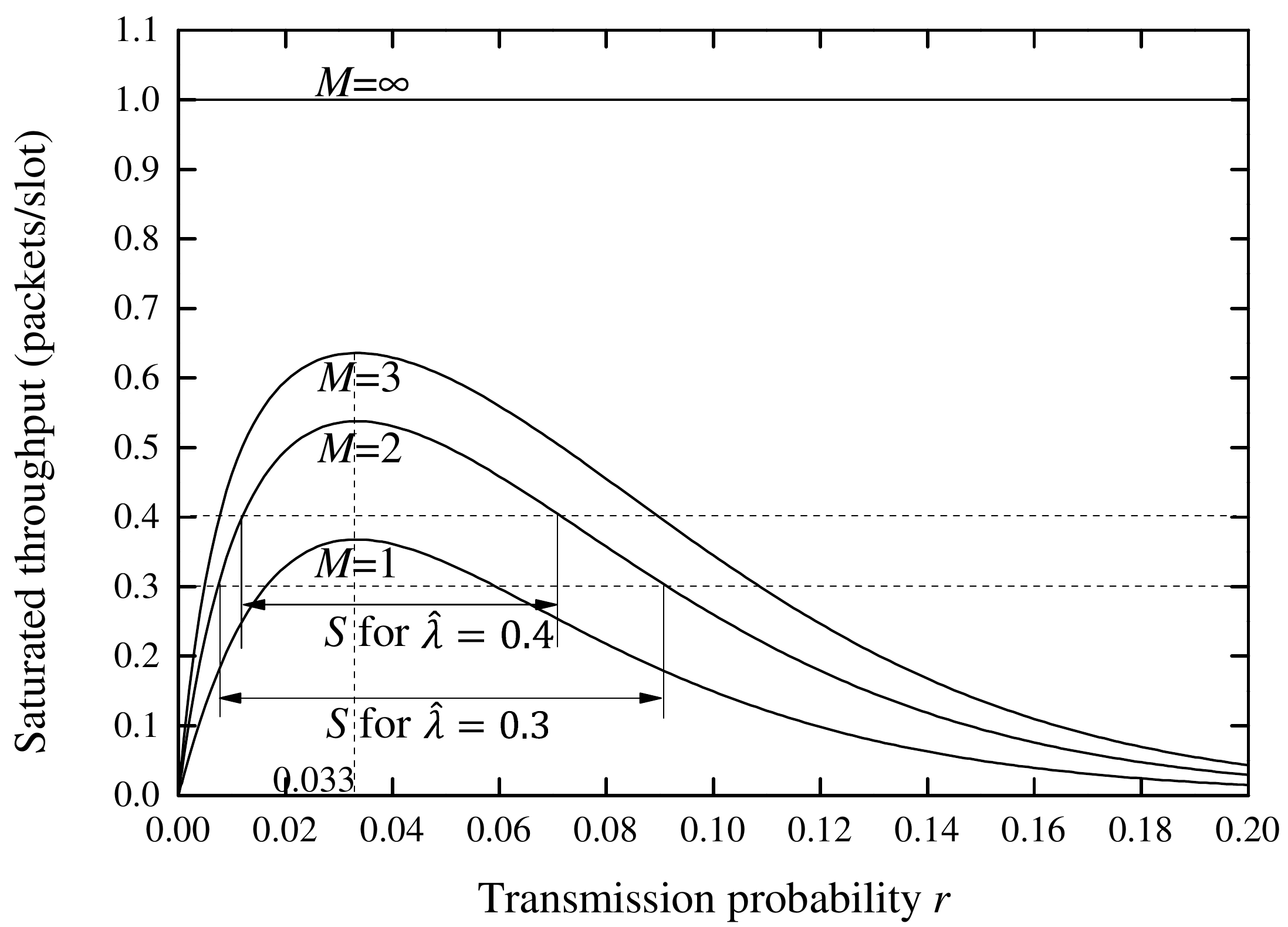}}
\subfigure[$n=50$]{
\label{fig3-b}
\includegraphics[scale=0.37]{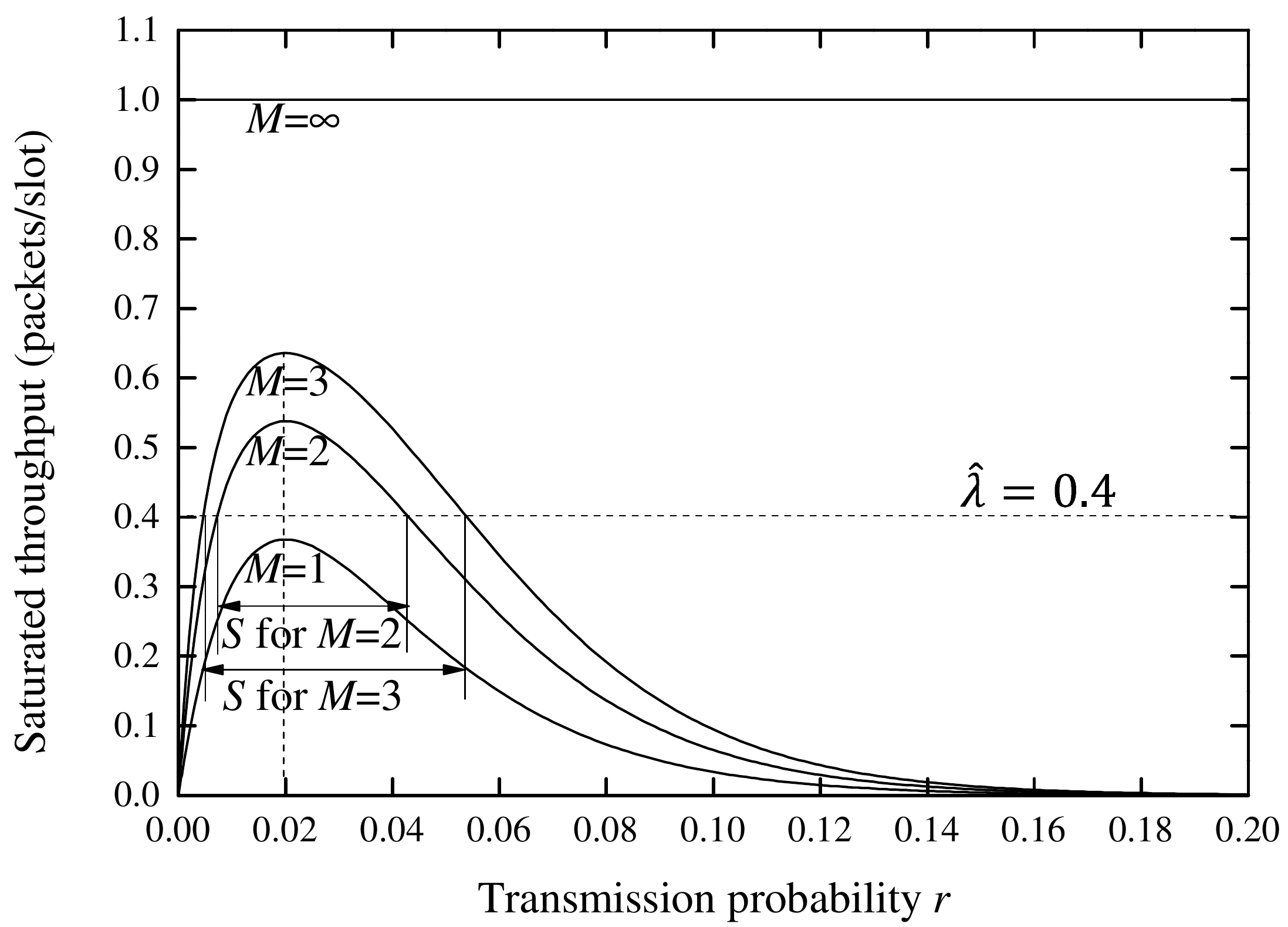}}
\caption{Saturated throughput versus transmission probability under different $M$s, where $n=30$ and $n=50$.}
\end{figure*}

To demonstrate this point, Fig. 3 plots the saturated throughput $\hat{\lambda}_{sat}$ changing with the transmission probability $r$. On one hand, given a finite $M$, $\hat{\lambda}_{sat}$ is a bell-shaped curve. In particular, $\hat{\lambda}_{sat}$ reaches its maximum value $M/(M+e-1)$ at $r=1/n$, (i.e., $G=nr=1$), which is quite similar to the case of the classical slotted Aloha. On the other hand, given $r$, $\hat{\lambda}_{sat}$ monotonically increases with $M$. The saturated throughput $\hat{\lambda}_{sat}=1$ for all $r$s in both Fig.~\ref{fig3-a} and \ref{fig3-b}, when $M$ increases to infinity.

\subsection{Stable Throughput Region}
However, the network in practice is unsaturated. It is clear that the throughput in this case is the smaller one of the aggregate input traffic rate $\hat{\lambda}$ and the saturated throughput, i.e., $\hat{\lambda}_{out}=min\{\hat{\lambda},\hat{\lambda}_{sat}\}$. If $\hat{\lambda}{\leq}\hat{\lambda}_{sat}$, the network has a stable throughput, i.e., $\hat{\lambda}_{out}=\hat{\lambda}$, which yields the stable throughput region as follows.

\newtheorem{myTheorem}{Theorem}
\begin{myTheorem}
Given the batch size $M$, and the number of nodes $n$, for any aggregate input traffic rate $\hat{\lambda}{\leq}\frac{M}{M+e-1}$, the transmission probability $r$ should be tuned in the following region
\begin{small}
\begin{equation}
r{\in}S=(0,1)\cap\left[\frac{-\mathbb{W}_{0}\left(-\frac{\hat{\lambda}}{M\left(1-\hat{\lambda}\right)+\hat{\lambda}}\right)}{n},\frac{-\mathbb{W}_{-1}\left(-\frac{\hat{\lambda}}{M\left(1-\hat{\lambda}\right)+\hat{\lambda}}\right)}{n}\right]
\label{Theorem1}
\end{equation}
\end{small}
to make the network stable, where $\mathbb{W}_{0}(\cdot)$ and $\mathbb{W}_{-1}(\cdot)$ are two principal branches of Lambert W function \cite{lambertw}.
\end{myTheorem}
\begin{IEEEproof}
Equation (\ref{Theorem1}) can be derived immediately by substituting (\ref{sat-throughput}) to $\hat{\lambda}{\leq}\hat{\lambda}_{sat}$.
\end{IEEEproof}

When $M$ is finite, it can be seen from (\ref{Theorem1}) that the stable throughput region $S$ shrinks with the increase of $n$ and $\hat{\lambda}$. On one hand, if $n$ increases, both $\frac{-\mathbb{W}_{0}\left(-\frac{\hat{\lambda}}{M\left(1-\hat{\lambda}\right)+\hat{\lambda}}\right)}{n}$ and $\frac{-\mathbb{W}_{-1}\left(-\frac{\hat{\lambda}}{M\left(1-\hat{\lambda}\right)+\hat{\lambda}}\right)}{n}$ decrease such that $S$ shrinks. When $n$ increases to infinity, $S$ disappears and the system becomes inherently unstable. For example, for $\hat{\lambda}=0.4$ packets/slot and $M=2$, $S$ in Fig.~\ref{fig3-b} when $n=50$ is smaller than that in Fig.~\ref{fig3-a} when $n=30$. On the other hand, with the increase of $\hat{\lambda}$, $-\mathbb{W}_{0}\left(-\frac{\hat{\lambda}}{M\left(1-\hat{\lambda}\right)+\hat{\lambda}}\right)$ increases and $-\mathbb{W}_{-1}\left(-\frac{\hat{\lambda}}{M\left(1-\hat{\lambda}\right)+\hat{\lambda}}\right)$ decreases such that $S$ shrinks also. In particular, once $\hat{\lambda}>M/(M+e-1)$, $S$ becomes empty and the system is unstable. As Fig.~\ref{fig3-a} displays, given $n=30$ and $M=2$, $S$ becomes empty when $\hat{\lambda}=0.6>2/(2+e-1)=0.538¦Ë$ packets/slot.

However, with the increase of $M$, $-\mathbb{W}_{0}\left(-\frac{\hat{\lambda}}{M\left(1-\hat{\lambda}\right)+\hat{\lambda}}\right)$ decreases while $-\mathbb{W}_{-1}\left(-\frac{\hat{\lambda}}{M\left(1-\hat{\lambda}\right)+\hat{\lambda}}\right)$ increases such that the stable throughput region $S$ becomes large. As Fig.~\ref{fig3-b} displays, for $n=50$ and $\hat{\lambda}=0.4$ packets/slot, the region $S$ when $M=3$ is larger than that when $M=2$. Especially, when $M$ increases to infinity, $S$ becomes (0,1) no matter what $n$ and $\hat{\lambda}$ are, because the saturated channel throughout $\hat{\lambda}_{sat}$ can reach $100\%$ for any $r\in(0,1)$ as we explain in Section II-A. This indicates that, in terms of network throughput, the batch-service slotted Aloha with $M=\infty$ is robust with respect to $r$.

Though the above model characterizes the behavior of the channel very well, it cannot delineate the queueing process of packets in each node and thus cannot obtain the delay performance. In practice, mean delay is an important criterion to measure network performance. The user's experience will be good if the mean delay is small. Hence, we study delay performance in the following sections.

\section{Vacation Model for an Access Node}
\begin{figure}[!t]
\centering
\includegraphics[scale=0.28]{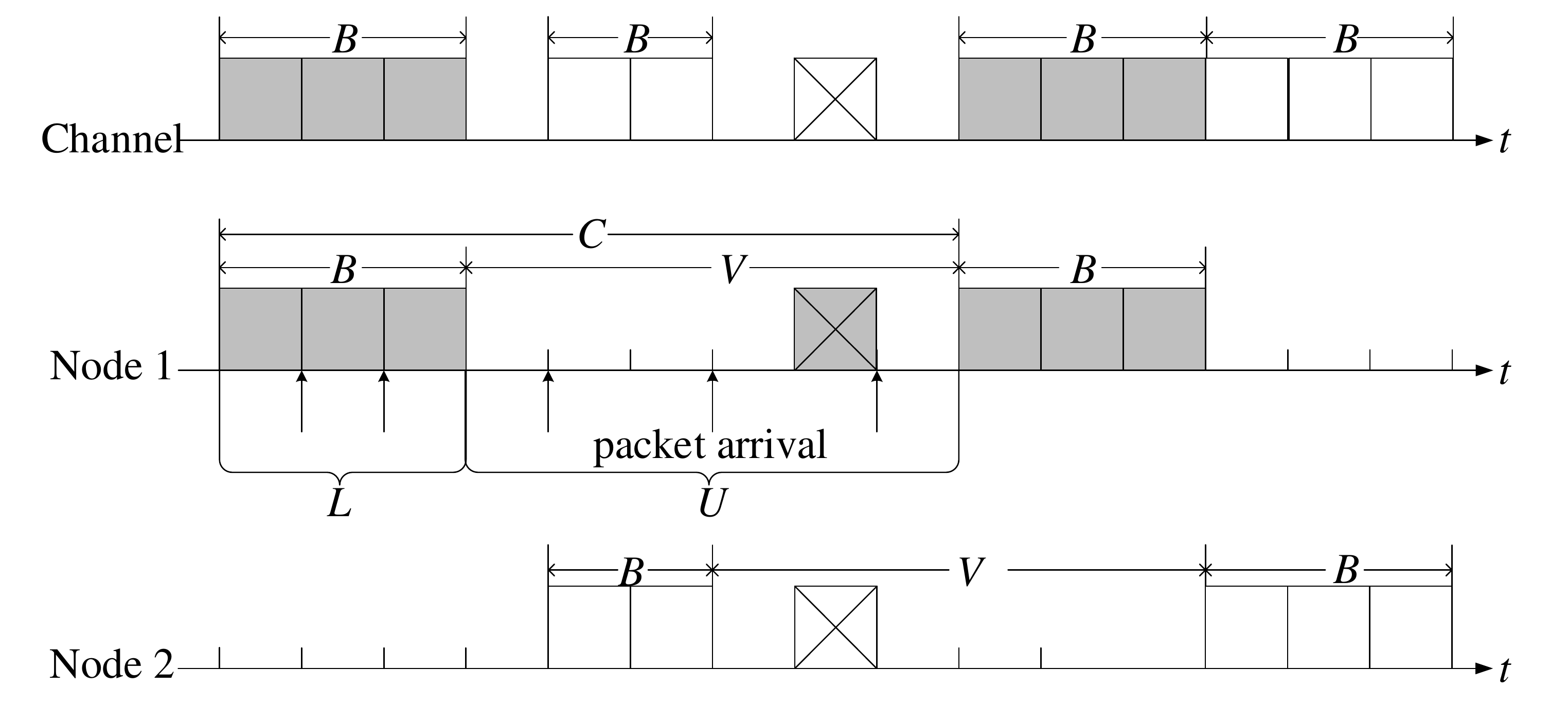}
\caption{Packet arrival and transmission process of slotted Aloha with batch service where $n=2$ and $M=3$.}
\label{fig4}
\end{figure}

In the slotted Aloha with batch service, the working process of each node can also be divided into cycles, each of which is defined as the duration between two consecutive successful attempts of the node. As Fig.~\ref{fig4} shows, a cycle, denoted by $C$, is composed of a busy period, denoted by $B$, and a vacation period, denoted by $V$. At the beginning of the slot in which a node makes a successful attempt, this node starts a busy period, during which this node sends out all the packets inside the virtual gate. After the busy period, the node releases the channel and a vacation period begins. This node competes for the channel immediately after it releases the channel if its buffer is not empty; otherwise, this node does that only after a new packet arrives. Thus, the vacation period is influenced by the packet arrival process. Also, the vacation period goes on until the node succeeds in channel competition again. If this node can make a successful attempt with high probability, the vacation period tends to be short. As we show in Section II-A, the successful probability of the node is determined by the attempt rate $G$. Thus, the vacation period is also governed by $G$. Therefore, with assumption A4 and A6, each node can be regarded as a Geo/D/1 queue with the vacation period governed by the packet arrival process and the attempt rate.

As \cite{pan2017TON} shows, the key to the analysis of a queueing system with vacation period governed by the packet arrival process is the distribution of the number of arrivals during the vacation period. In the slotted Aloha with batch service, a vacation period of a node may include multiple busy periods of other nodes. Fig.~\ref{fig4} illustrates such an example, where a vacation period of node 1 includes one busy period of node 2 since node 2 succeeds in the 5th slot before node 1 succeeds again in the 10th slot. Thus, the distribution of the number of arrivals during a vacation period depends on that during the busy period. In the following, we derive these two distributions.

According to assumption A4, the number of arrivals during a busy period, denoted by $L$, obeys a binomial distribution if the busy period is given. Define $b_{j}{\triangleq}Pr\{B=j\}$ as the probability that the busy period lasts for $j$ slots, where $j=1,2,\cdots,M$. The generating function of $b_j$ is $B(z){\triangleq}E[z^{B}]=\sum_{j=1}^{M}b_{j}z^{j}$. Thus, the probability generating function of the number of arrivals during the busy period is given by
\begin{equation}
\begin{split}
L(z)&=E[z^L]   \\
    &=\sum_{j=1}^{M}E[z^L|B=j]b_j  \\
    &=\sum_{j=1}^{M}b_j\left[\sum_{k=0}^{j}\binom{j}{k}{\lambda}^k(1-\lambda)^{j-k}z^k\right] \\
    &=\sum_{j=1}^{M}b_j(1-\lambda+\lambda{z})^j  \\
    &=B(1-\lambda+\lambda{z})  \\
    &=1+{\lambda}B^{'}(1)(z-1)+\frac{{\lambda}^2}{2}B^{''}(1)(z-1)^2+\cdots.
\label{gen-L}
\end{split}
\end{equation}
Given the aggregate input traffic rate $\hat{\lambda}$, the input traffic rate of each node $\lambda=\frac{\hat{\lambda}}{n}$ is on the order of $o(\frac{1}{n})$. Thus, $L(z)$ can be expressed as
\begin{equation}
L(z)=1-\lambda\overline{B}+\lambda\overline{B}z+o\left(\frac{1}{n}\right),
\label{gen-L-2}
\end{equation}
where $\overline{B}=B^{'}(1)$ is the mean busy period. Equation (\ref{gen-L-2}) clearly indicates that the distribution of $L$ approaches a Bernoulli distribution with the mean of $\lambda\overline{B}$, when the number of nodes $n$ is sufficiently large. This result is consistent with intuition. When $n$ is sufficiently large, $\lambda$ is so small that the probability that more than one packet arrives at the node during the busy period is negligible. Also, in a stable network, the system throughput $\hat{\lambda}_{out}$ equals the aggregate input traffic rate $\hat{\lambda}$, and thus $\overline{B}$ can be determined according to (\ref{throughput}) as follows:
\begin{equation}
\overline{B}=\frac{\hat{\lambda}}{1-\hat{\lambda}}\left(\frac{1}{Ge^{-G}}-1\right).
\label{meanB}
\end{equation}

\begin{figure}[!t]
\centering
\includegraphics[scale=0.35]{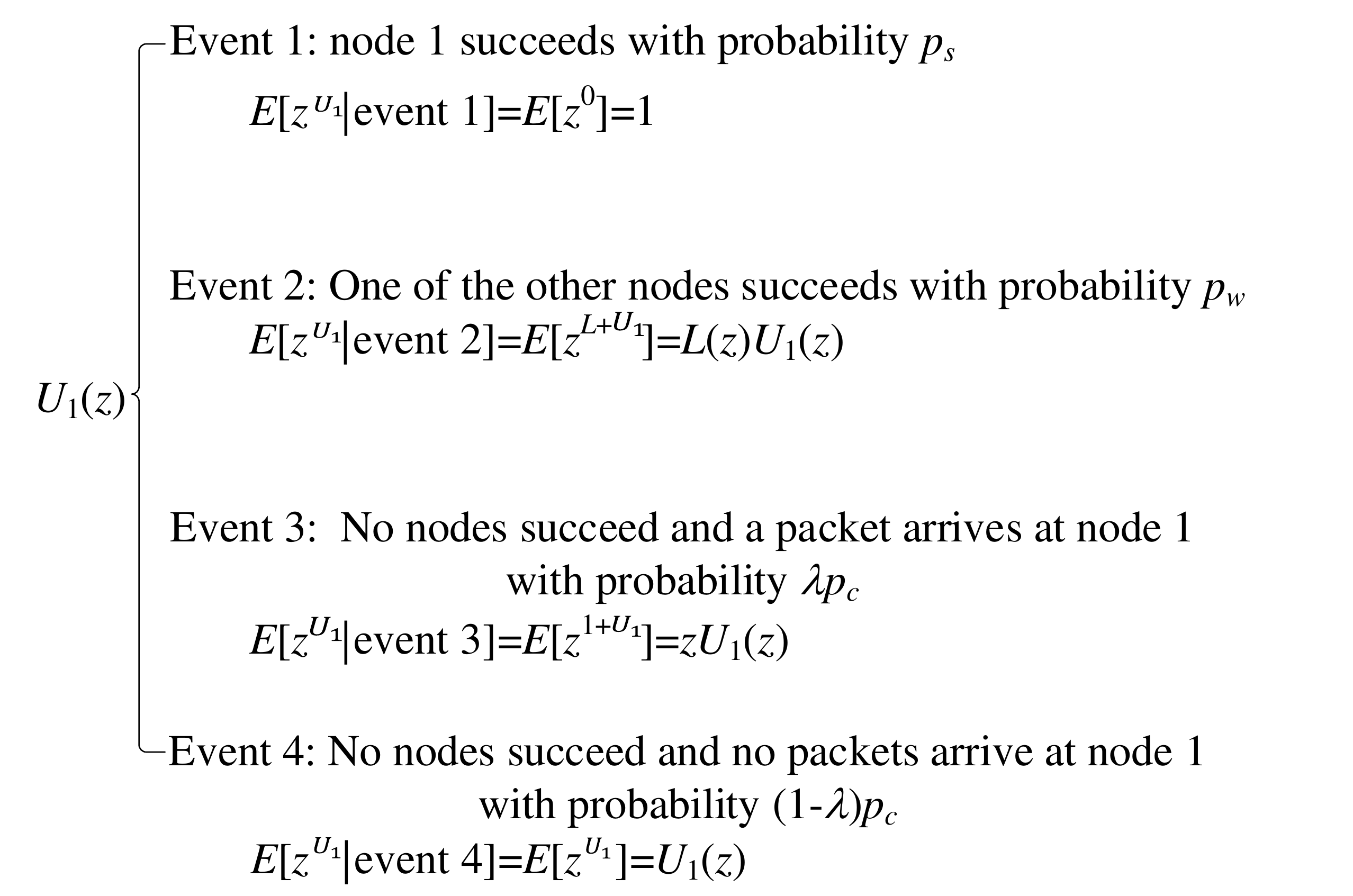}
\caption{Renewal process of the number of arrivals $U_1$ during vacation period $Y_1$.}
\label{fig5}
\end{figure}

When the busy period finishes, the node starts a vacation period. The buffer at this epoch may be empty or not. If the buffer is not empty, the node competes for the channel immediately. Let $Y_1$ be the vacation period and $U_1$ be the number of arrivals during $Y_1$ in this case. As Fig.~\ref{fig5} shows, there are four events that may occur in the first slot of the vacation period.
\begin{enumerate}[(a)]
\item
If node 1 makes an attempt with probability $r$ while the other $n-1$ nodes do not attempt with probability $e^{-G}$, it will begin another busy period immediately. Thus, this event happens with probability $p_s=re^{-G}$. In this case, both the vacation period and the number of arrivals during this vacation are zero, that is $U_1=0$.
\item
If node 1 doesn't make an attempt with probability $1-r$ and one of the other $n-1$ nodes transmits a packet with probability $Ge^{-G}$, this successful node will start a busy period immediately. Thus, this event happens with probability $p_w=(1-r)Ge^{-G}$. In this case, node 1 will restart the channel competition after this busy period. Recall that the number of arrivals during a busy period is $L$. The number of arrivals during the vacation is equal to $L+U_1$, because of the memoryless property of the channel competition process.
\item
If no node or multiple nodes make attempts, which occurs with probability $p_c=1-p_s-p_w$, no one can succeed in this slot and node 1 will compete for the channel in the next slot. In this case, if a packet arrives at node 1 with probability $\lambda$ in the current slot, the number of arrivals during the vacation period is now equal to $1+U_1$.
\item
If no node succeeds with probability $p_c$ and no packet arrives at node 1 with probability $1-\lambda$, the number of arrivals during the vacation period will be equal to $U_1$.
\end{enumerate}
The above clearly indicates that the channel access procedure is a renewal process. Conditioning on the event that occurs in the first slot, the probability generating function of $U_1$ satisfies the following equation:
\begin{small}
\begin{equation*}
U_{1}(z)=E[z^{U_1}]=p_s+p_wL(z)U_{1}(z)+{\lambda}p_czU_{1}(z)+(1-\lambda)p_cU_{1}(z),
\end{equation*}
\end{small}
which yields,
\begin{equation}
U_{1}(z)=\frac{p_s}{1-p_c(1-\lambda+{\lambda}z)-p_wL(z)}
\label{gen-U1}
\end{equation}

\begin{figure}[!t]
\centering
\includegraphics[scale=0.35]{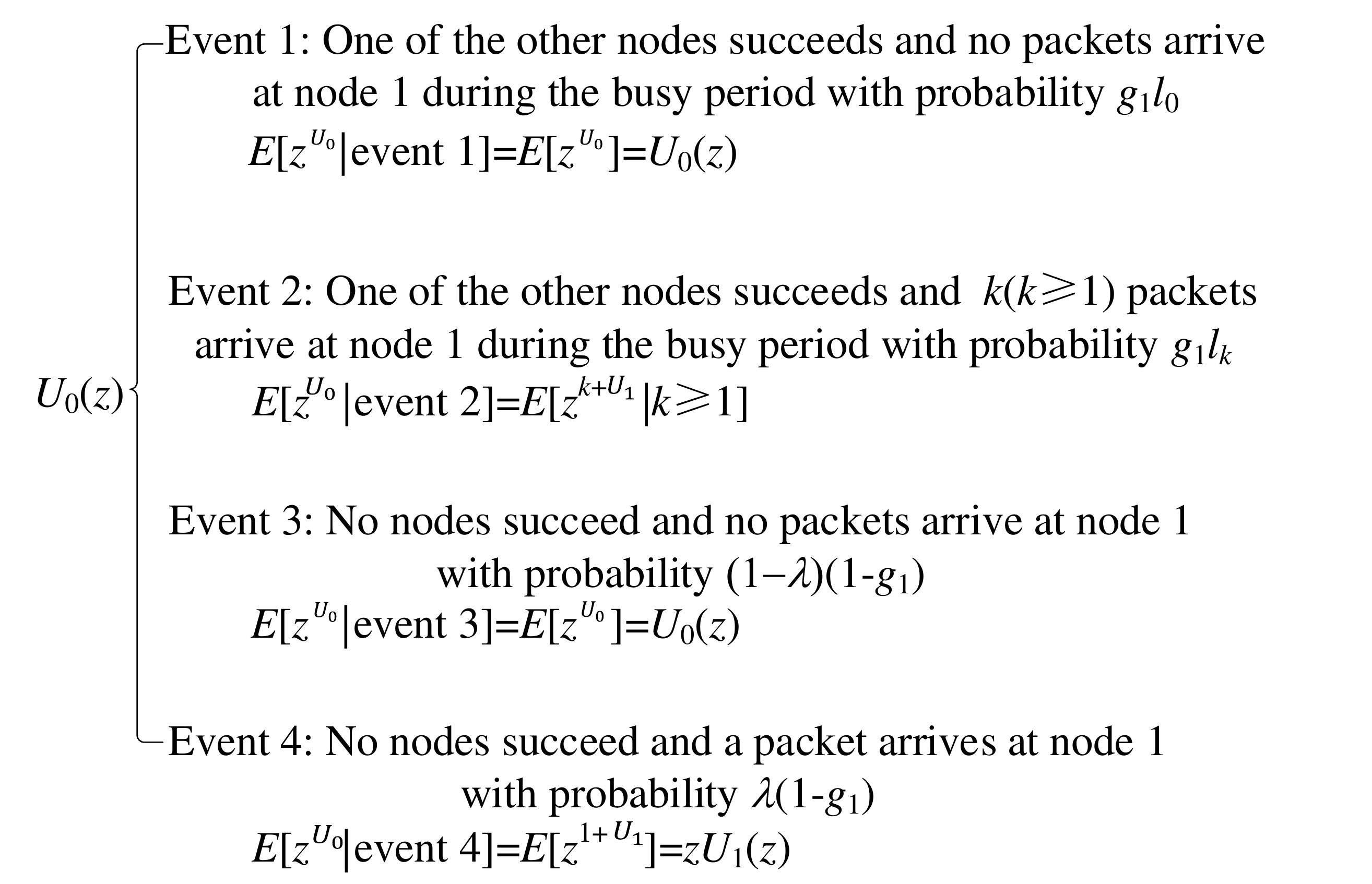}
\caption{Renewal process of the number of arrivals $U_0$ during vacation period $Y_0$.}
\label{fig6}
\end{figure}

On the other hand, if the buffer is empty when its busy period finishes, the node competes for the channel only after a new packet arrives. Let $Y_0$ be the vacation period and $U_0$ be the number of arrivals during $Y_0$ in this case. As Fig.~\ref{fig6} shows, four events may happen in the first slot of the vacation period, which is quite similar to the case in Fig.~\ref{fig5}. Thus, the derivation of $U_0(z){\triangleq}E[z^{U_0}]$ is similar to that of $U_1(z)$, and is omitted here. Instead, we only give the expression of $U_0(z)$ as follows:
\begin{equation}
U_0(z)=\frac{\lambda(1-g_1)z+g_1[L(z)-l_0]}{1-(1-\lambda)(1-g_1)-g_1l_0}U_1(z),
\label{gen-U0}
\end{equation}
where $g_1=Ge^{-G}$ and $l_0{\triangleq}Pr\{L=0\}=L(0)=B(1-\lambda)$ is the probability that no packet arrives in a busy period.

The channel competition process of one node can be viewed as a series of Bernoulli trials, which are terminated once the node makes a successful trial. The number of trials needed by a node for success is geometrically distributed. If a trial is not successful, two events may occur: 1) no one succeeds, and the sojourn time of this event is one slot and the number of arrivals in a slot is a Bernoulli variable with rate $\lambda$; 2) one of the other nodes succeeds, and the sojourn time of this event is a busy period $B$ and the number of arrivals in a busy period is a Bernoulli variable with rate $\lambda\overline{B}$ if the number of nodes $n$ is large enough. In other words, the number of arrivals during a vacation period is the sum of two kinds of Bernoulli random variables if $n$ is sufficiently large, and the number of the Bernoulli random variables is a geometric random variable. This hints that the number of arrivals during a vacation period may be a geometrically distributed random variable.

Let $U$ be the number of arrivals during the vacation period and $U(z)$ be the probability generating function of $U$. We have the following result.
\newtheorem{myLemma}{Lemma}
\begin{myLemma}
When the number of nodes $n$ is sufficiently large, $U(z)$ can be expressed as
\begin{small}
\[ U(z)=\begin{cases}
U_1(z), & \textrm{if buffer is non-empty at the start of a vacation} \\
U_0(z), & \textrm{otherwise}
\end{cases} \]
\end{small}
where
\begin{equation}
U_1(z)=\frac{\beta}{1-(1-\beta)z}+o\left(\frac{1}{n}\right),
\label{U1}
\end{equation}
\begin{equation}
U_0(z)=zU_1(z)+o\left(\frac{1}{n}\right)=\frac{{\beta}z}{1-(1-\beta)z}+o\left(\frac{1}{n}\right),
\label{U0}
\end{equation}
and
\begin{small}
\begin{equation}
\begin{split}
\beta &=\frac{p_s}{p_s+{\lambda}p_c+{\lambda}\overline{B}p_w} \\
      &=\frac{(1-\hat{\lambda})re^{-G}}{(1-\hat{\lambda})re^{-G}+\lambda[(1-\hat{\lambda})(1-re^{-G})+(1-r)(\hat{\lambda}-Ge^{-G})]}.
\end{split}
\end{equation}
\end{small}
\end{myLemma}
\begin{IEEEproof}
It is clear that $U=U_1$ if the buffer is not empty at the start of the vacation period; otherwise, $U=U_0$. When $n$ is sufficiently large, the number of arrivals during a busy period $L$ approaches a Bernoulli random variable, as (\ref{gen-L-2}) shows. Substituting (\ref{gen-L-2}) into (\ref{gen-U1}), we obtain
\begin{equation*}
\begin{split}
U_1(z)&=\frac{p_s}{1-p_c(1-\lambda+{\lambda}z)-p_w(1-\lambda\overline{B}+\lambda\overline{B}z)}+o\left(\frac{1}{n}\right) \\
      &=\frac{\frac{p_s}{p_s+{\lambda}p_c+\lambda\overline{B}p_w}}{1-\left(1-\frac{p_s}{p_s+{\lambda}p_c+\lambda\overline{B}p_w}\right)z}+o\left(\frac{1}{n}\right).
\end{split}
\end{equation*}
Similarly, substituting (\ref{gen-L-2}) and $l_0=B(1-\lambda)=1-\lambda\overline{B}+o(\frac{1}{n})$ into (\ref{gen-U0}) produces $U_0(z)$ in (\ref{U0}).
\end{IEEEproof}

It follows from Lemma 1 that $\overline{U_0}{\triangleq}U_0^{'}(1)=1+U_1^{'}(1){\triangleq}1+\overline{U_1}$. Once the average number of arrivals during the vacation period is obtained, we can immediately derive the average vacation periods as follows:
\begin{small}
\begin{equation}
\overline{Y_1}=\frac{U_1^{'}(1)}{\lambda}=\frac{(1-\hat{\lambda})(1-re^{-G})+(1-r)(\hat{\lambda}-Ge^{-G})}{(1-\hat{\lambda})re^{-G}},
\label{mean-Y1}
\end{equation}
\end{small}
and
\begin{equation}
\overline{Y_0}=\frac{U_0^{'}(1)}{\lambda}=\frac{1}{\lambda}+\overline{Y_1}.
\label{mean-Y0}
\end{equation}

As we mention at the beginning of this section, the vacation period is also influenced by the attempt rate $G$, which can be confirmed by (\ref{mean-Y1}) and (\ref{mean-Y0}). We thus analyze $G$ in the next section.

\section{Attempt Rate}
Once the channel becomes free, each backlogged node makes an attempt in a slot with probability $r$. Thus, the attempt rate is given by
\begin{equation}
G=np_{ne}r,
\label{G-r}
\end{equation}
where $p_{ne}$ is the probability that the buffer of a node is non-empty or a node is backlogged.

\begin{figure}[!t]
\centering
\includegraphics[scale=0.5]{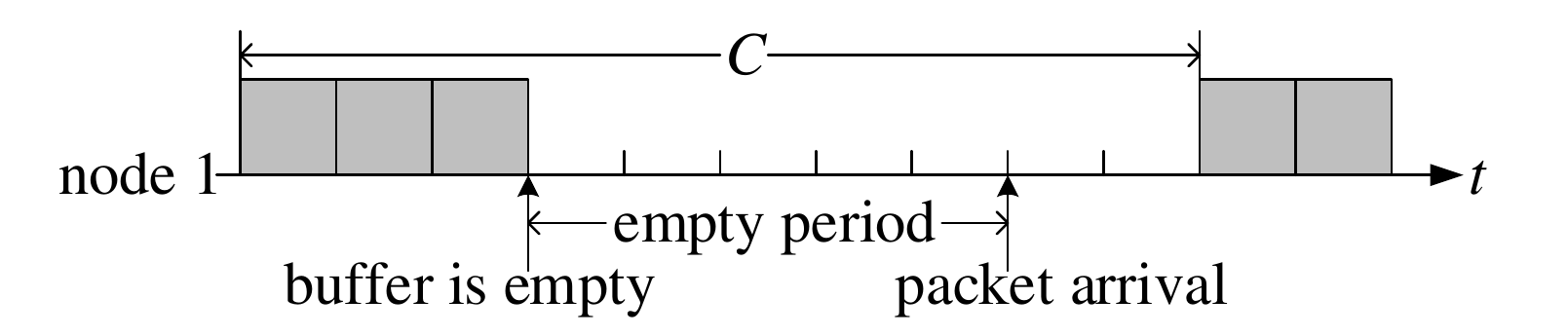}
\caption{Empty period of a node in a cycle.}
\label{fig7}
\end{figure}

As Fig.~\ref{fig7} depicts, the buffer of a node may become empty only at the end of a busy period for this node. The empty state terminates when a new packet arrives. The average inter-arrival time of the packets is $1/\lambda$. Let $p_0$ be the probability that the buffer of a node is empty at the end of a busy period. Thus, the average time that the buffer of a node stays empty in a cycle is $p_0/\lambda$, and thus the probability that the buffer of a node is empty is given by
\begin{equation}
1-p_{ne}=\frac{p_0/\lambda}{\overline{C}}.
\label{em-pro}
\end{equation}
In a stable network, the mean busy period and the mean cycle time have the following relationship:
\begin{equation}
\overline{B}=\lambda\overline{C}.
\label{B-C}
\end{equation}
Combining (\ref{em-pro}) and (\ref{B-C}), we obtain
\begin{equation*}
p_{ne}=1-\frac{p_0}{\overline{B}},
\end{equation*}
which yields
\begin{equation}
G=np_{ne}r=nr\left(1-\frac{p_0}{\overline{B}}\right),
\label{G-p0}
\end{equation}
where $\overline{B}$ is given in (\ref{meanB}). Thus, the buffer empty probability at the end of a busy period of this node $p_0$ is the key to derive the attempt rate $G$.

Let $Q_t$ and $P_t$ be the queue lengths at the start and the end of the $t$-th busy period of a node, respectively. Define $q_k{\triangleq}\lim_{t{\to}\infty}Pr\{Q_t=k\}$ and $p_k{\triangleq}\lim_{t{\to}\infty}Pr\{P_t=k\}$, and let $Q(z)$ and $P(z)$ be the generating functions of $q_k$ and $p_k$, where $k=0,1,{\cdots}$.

\begin{figure*}[!t]
\centering
\includegraphics[scale=0.4]{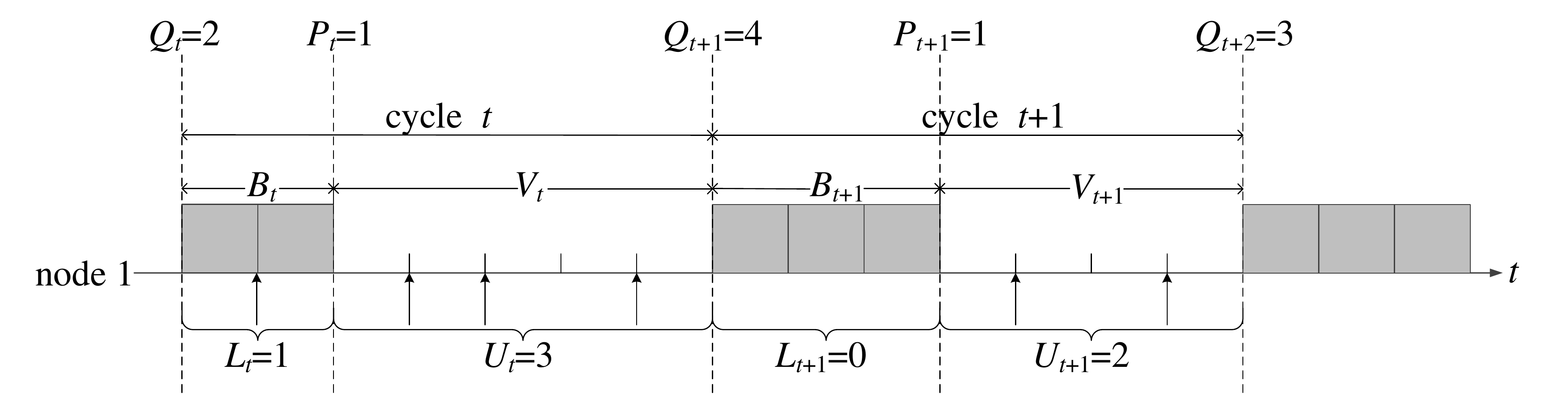}
\caption{Packet transmission process of node 1 where $M=3$.}
\label{fig8}
\end{figure*}

On one hand, $p_k$ depends on $q_k$. As Fig.~\ref{fig8} shows, according to the batch service discipline, if $Q_t$ is less than $M$, all the $Q_t$ packets will be transmitted in the busy period of cycle $t$; otherwise, only the first $M$ packets will be sent. Thus, the number of packets served in the $t$-th busy period, denoted by $B_t$, is determined by
\begin{equation}
B_t=Q_t-(Q_t-M)^{+},
\label{Bt}
\end{equation}
where $x^{+}{\triangleq}max\{x,0\}$. Let $L_t$ denote the number of packets that arrive during $B_t$. The queue length at the end of the $t$-th busy period $P_t$ can be given by
\begin{equation}
P_t=Q_t-B_t+L_t=(Q_t-M)^{+}+L_t.
\label{Pt}
\end{equation}
In the steady state, we have
\begin{equation}
P=\lim_{t{\to}\infty}P_t=(Q-M)^{+}+L.
\label{P}
\end{equation}
Therefore, the probability generating function of $P$ is
\begin{equation}
\begin{split}
P(z){\triangleq}&E[z^P] \\
    =&E[z^{(Q-M)^{+}+L}]   \\
    =&\sum_{k=1}^{\infty}E[z^{(Q-M)^{+}+L}|Q=k,B=k-(k-M)^{+}]q_k  \\
    =&\sum_{k=1}^{M-1}E[z^L|B=k]q_k+\sum_{k=M}^{\infty}z^{k-M}E[z^L|B=M]q_k  \\
    =&\sum_{k=1}^{M-1}(1-\lambda+\lambda{z})^kq_k+\sum_{k=M}^{\infty}z^{k-M}(1-\lambda+\lambda{z})^Mq_k.
\label{Pz}
\end{split}
\end{equation}
Hence, the probability that the buffer of a node is empty at the end of a busy period is
\begin{equation}
p_0=P(0)=\sum_{k=1}^{M}(1-\lambda)^kq_k,
\label{p0}
\end{equation}
Equation (\ref{Pz}) clearly indicates that the distribution of the queue length at the end of the busy period $p_k$ depends on that of the queue length at the start of the busy period $q_k$.

\begin{figure*}[!t]
\centering
\subfigure[$M=1$]{
\label{fig9-a}
\includegraphics[scale=0.23]{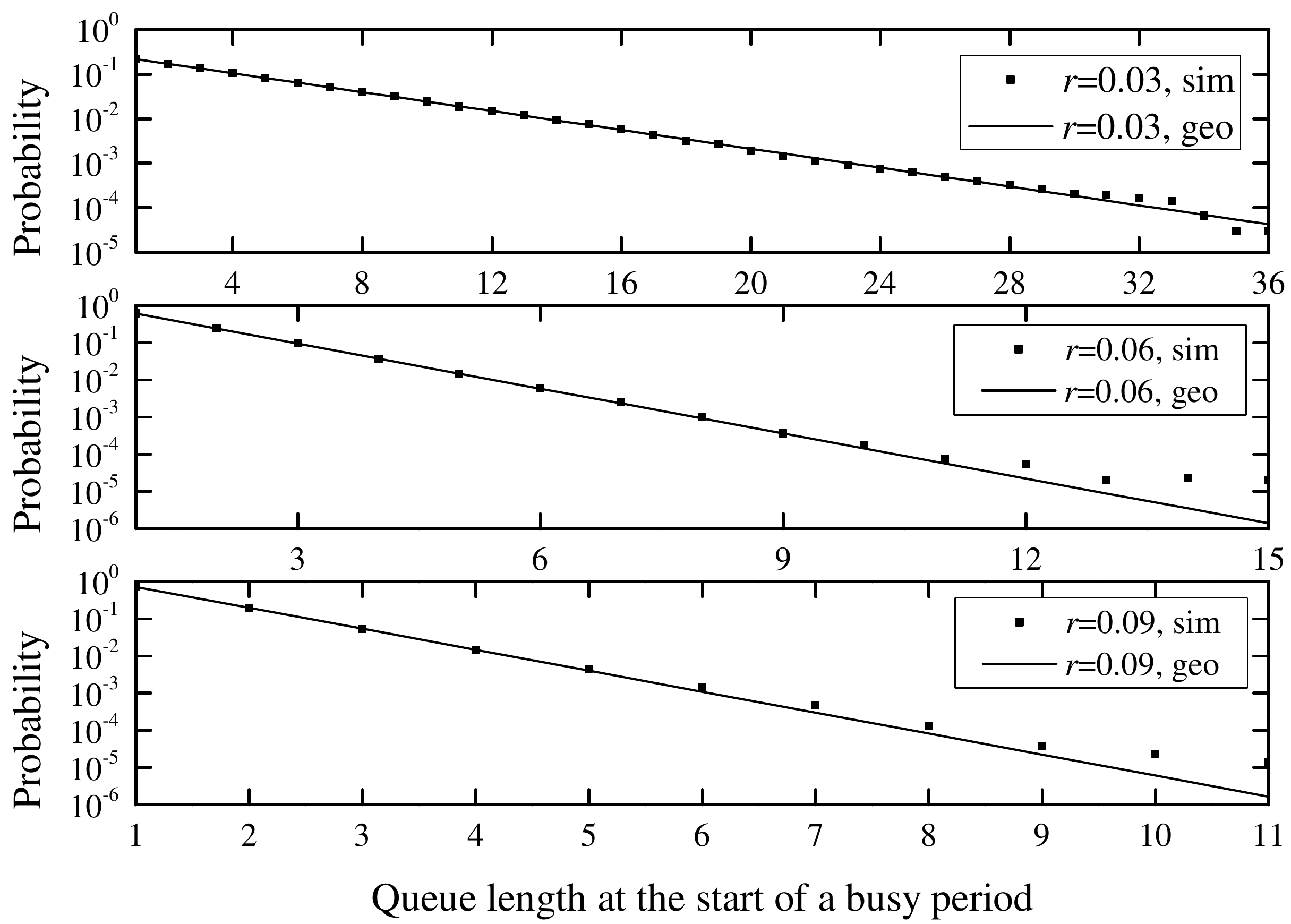}}
\subfigure[$M=10$]{
\label{fig9-b}
\includegraphics[scale=0.23]{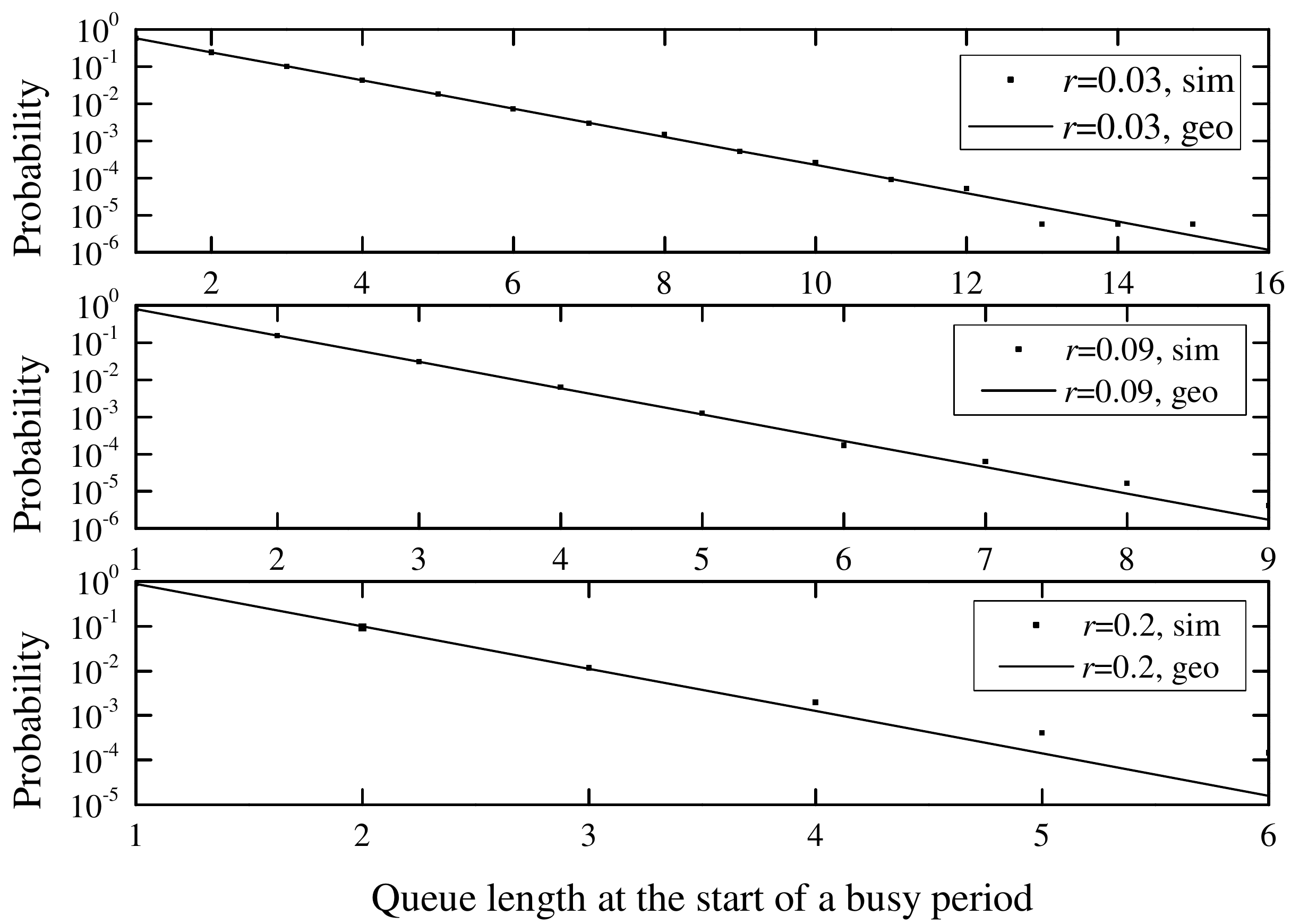}}
\subfigure[$M=\infty$]{
\label{fig9-c}
\includegraphics[scale=0.23]{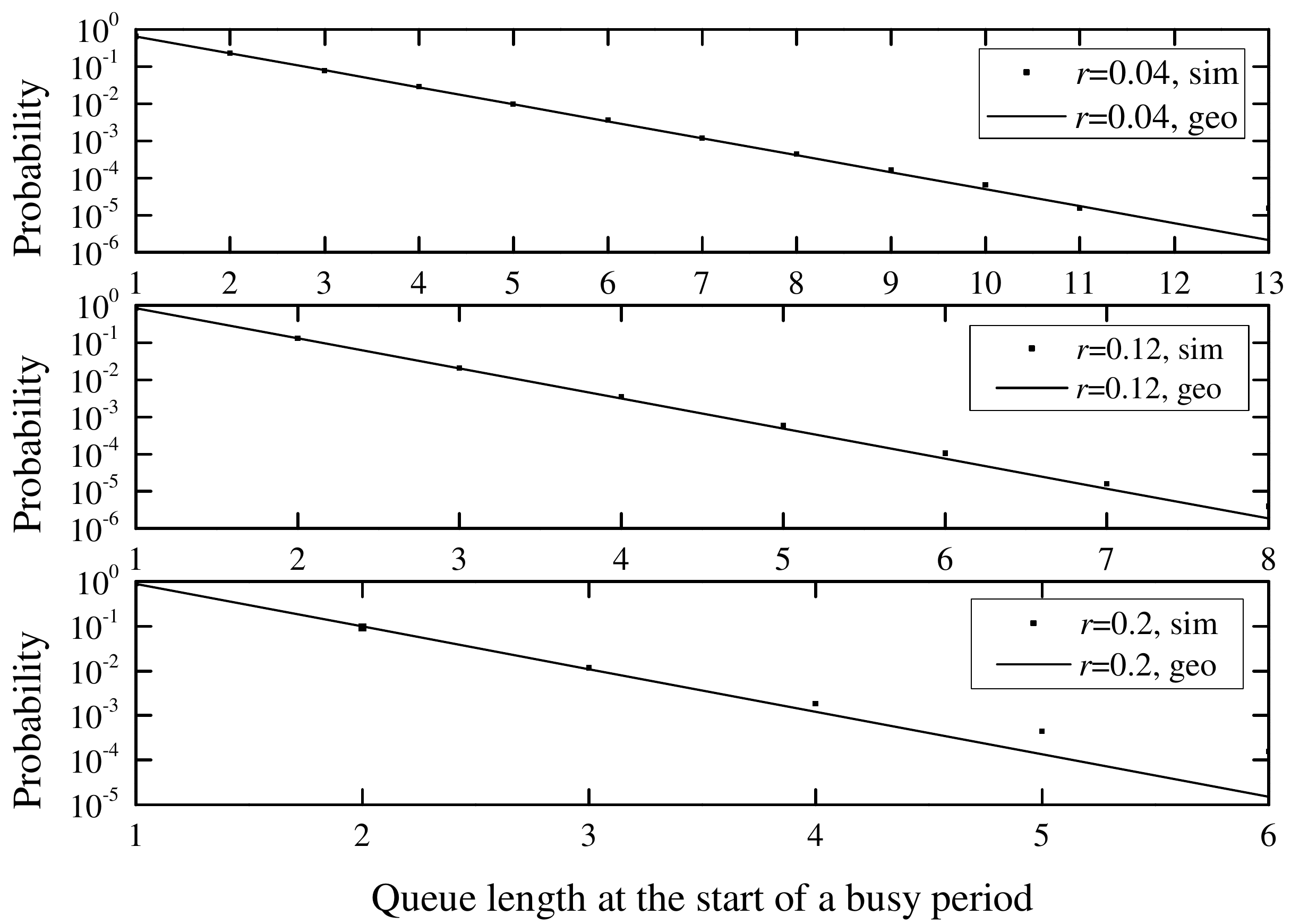}}
\caption{Distribution $q_k$ under different $M$s and $r$s, where $n=20$ and $\hat{\lambda}=0.3$ packets/slot.}
\end{figure*}

On the other hand, $q_k$ is also determined by $p_k$. As Fig.~\ref{fig8} depicts, the queue length at the start of the $(t+1)$-th busy period $Q_{t+1}$ is composed of the packets waiting in the buffer at the end of the $t$-th busy period $P_t$ and the packets that arrive during the $t$-th vacation period, denoted by $U_t$, i.e.,
\begin{equation}
Q_{t+1}=P_{t}+U_{t}.
\label{Qt}
\end{equation}
In the steady state, there is
\begin{equation}
Q=\lim_{t{\to}\infty}=Q_{t+1}=P+U.
\label{Q}
\end{equation}
Accordingly, its probability generating function is given by
\begin{equation}
\begin{split}
Q(z){\triangleq}&E[z^Q]  \\
    =&E[z^{P+U}]   \\
    =&E[z^{P+U_0}|P=0]p_0+\sum_{k=1}^{\infty}E[z^{P+U_1}|P=k]p_k  \\
    =&p_0E[z^{U_0}]+\sum_{k=1}^{\infty}E[z^{U_1}]E[z^P|P=k]p_k    \\
    =&p_0U_0(z)+[P(z)-p_0]U_1(z).
\label{Qz}
\end{split}
\end{equation}
Equations (\ref{Pz}) and (\ref{Qz}) show that $P(z)$ and $Q(z)$ couple with each other and thus can only be solved numerically in general.

However, we will demonstrate that $P(z)$ and $Q(z)$ have closed-form solutions when the number of nodes $n$ is sufficiently large. Recall that Lemma 1 shows that the distribution of the number of arrivals during the vacation period $U$ approaches a geometric distribution when $n$ is sufficiently large. Also, the vacation period of a node is typically much longer than its busy period, and thus the number of arrivals during the vacation period predominates the number of packets in the buffer at the start of the next busy period. This hints that the distribution of the queue length at the start of the busy period $Q$ may also approach a geometric distribution.
\newtheorem{myLemma1}[myLemma]{Lemma}
\begin{myLemma1}
When $n$ is sufficiently large, $Q(z)$ can be expressed as
\begin{equation}
Q(z)=\frac{\alpha{z}}{1-(1-\alpha)z}+o\left(\frac{1}{n}\right),
\label{Qz-geo}
\end{equation}
where $\alpha$ is given by
\begin{equation}
\alpha=1-\frac{G}{nr}.
\label{alpha}
\end{equation}
\end{myLemma1}
\begin{IEEEproof}
See APPENDIX A.
\end{IEEEproof}

Fig. 9 verifies the analytical result in (\ref{Qz-geo}) via simulation when the number of nodes $n=20$ and the aggregate input traffic rate $\hat{\lambda}=n\lambda=0.3$ packets/slot. Fig. 9 clearly shows that $n=20$ is already large enough to ensure that the result in (\ref{Qz-geo}) is very accurate.

According to Lemma 2, we can easily obtain $q_k$ when $n$ is sufficiently large as follows:
\begin{equation}
q_k=\frac{1}{k!}\frac{d^kQ(z)}{dz^k}|_{z=0}=\alpha(1-\alpha)^{k-1}, k=1,2,\cdots
\label{qk}
\end{equation}
Substituting (\ref{qk}) into (\ref{p0}), the probability that the buffer of a node is empty at the end of a busy period is given by
\begin{equation}
\begin{split}
p_0=&\sum_{k=1}^M(1-\lambda)^k\alpha(1-\alpha)^{k-1}  \\
   =&\frac{(1-\lambda)\alpha[1-(1-\lambda)^M(1-\alpha)^M]}{1-(1-\lambda)(1-\alpha)}.
\label{p0-alpha}
\end{split}
\end{equation}
Given $\hat{\lambda}$, the input traffic rate of each node $\lambda=\hat{\lambda}/n$ is on the order of $o(\frac{1}{n})$. Thus, when $n$ is sufficiently large, $1-\lambda$ is approximately equal to 1. In this case, we have
\begin{equation}
p_0{\approx}1-(1-\alpha)^M=1-\left(\frac{G}{nr}\right)^M.
\label{p0-G}
\end{equation}
With $\overline{B}$ in (\ref{meanB}) and $p_0$ in (\ref{p0-G}), we are now ready to derive the attempt rate $G$ using (\ref{G-p0}) as follows.
\begin{figure*}[!t]
\centering
\includegraphics[scale=0.6]{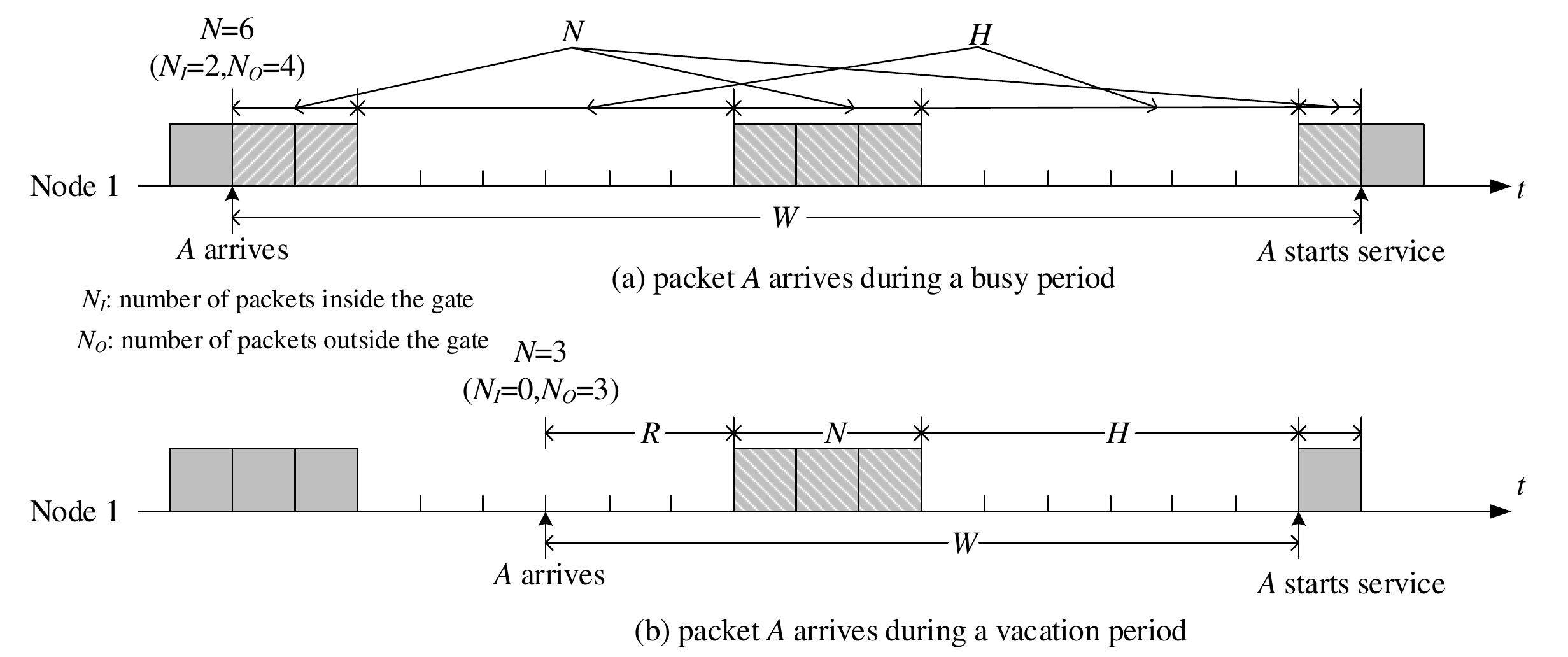}
\caption{Waiting time of packet $A$ where $M=3$.}
\label{fig10}
\end{figure*}

\newtheorem{myTheorem1}[myTheorem]{Theorem}
\begin{myTheorem1}
Given the aggregate input traffic rate $\hat{\lambda}{\leq}\frac{M}{M+e-1}$, and a transmission probability $r$ in the stable throughput region (\ref{Theorem1}), the attempt rate $G$ of the system with a sufficiently large $n$ is the solution of the following equation:
\begin{equation}
\frac{1-\left(\frac{G}{nr}\right)^M}{1-\frac{G}{nr}}=\frac{\hat{\lambda}}{1-\hat{\lambda}}\left(\frac{1}{Ge^{-G}}-1\right).
\label{Theorem2}
\end{equation}
\end{myTheorem1}
\begin{IEEEproof}
Substituting (\ref{meanB}) and (\ref{p0-G}) into (\ref{G-p0}), we can get Eq. (\ref{Theorem2}) after some reconfigurations.
\end{IEEEproof}
Furthermore, APPENDIX B shows how to solve $G$ from (\ref{Theorem2}).

\section{Delay Performance}

Consider a packet, denoted by $A$, arrives at a node, say node 1 in Fig.~\ref{fig10}, when there are $N$ packets waiting in the buffer. Packet $A$ may arrive at node 1 during a busy period or a vacation period. If packet $A$ arrives during a busy period as Fig. 10(a) plots, the first of the $N$ packets in the buffer gets service immediately; otherwise, as Fig. 10(b) shows, packet transmission of node 1 starts after the vacation finishes. This implies that packet $A$ may wait for a residual vacation period. Also, packet $A$ has to wait for the service completion of the $N$ packets that stay ahead of it in the queue. Furthermore, due to the batch size, it may take node 1 several busy periods to send out all of the $N$ packets. In this case, packet $A$ has to wait for multiple vacation periods in addition. In summary, the waiting time of packet $A$ consists of the following three components:
\begin{enumerate}[(a)]
\item
The residual vacation period seen by packet $A$, which is denoted by $R$;
\item
The total transmission time of $N$ packets that are waiting in the queue upon the arrival of packet $A$;
\item
The total complete vacation periods experienced by packet $A$ before transmission, which is denoted by $H$.
\end{enumerate}
Thus, the mean waiting time of the packet, denoted by $\overline{W}$, can be expressed as
\begin{equation}
\overline{W}=\overline{R}+\overline{N}+\overline{H},
\label{meanW}
\end{equation}
where $\overline{N}=\lambda\overline{W}$ according to Little's law. In the following, we derive $\overline{R}$ and $\overline{H}$ to complete the derivation.

As Section \uppercase\expandafter{\romannumeral2} mentions, the vacation periods of a node can be divided to two types, depending on whether the buffer of this node is empty or not. Let $\xi$ be the type of vacation period, during which packet $A$ arrives
\[ \xi=\begin{cases}
0, &\textrm{packet A arrives during a vacation period $Y_0$}  \\
1, &\textrm{packet A arrives during a vacation period $Y_1$}.
\end{cases} \]
Recall that the vacation periods $Y_0$ and $Y_1$ occur with probability $p_0$ and probability $1-p_0$, and the mean numbers of arrivals during $Y_0$ and $Y_1$ are $U_0^{'}(1)$ and $U_1^{'}(1)$, respectively. Let $Pr\{\xi=0\}$ and $Pr\{\xi=1\}$ be the probabilities that packet $A$ arrives during vacation period $Y_0$ and that packet $A$ arrives during vacation period $Y_1$. We have
\begin{small}
\begin{equation}
\begin{split}
Pr\{\xi=0\}=&Pr\{\xi=0|\textrm{$A$ arrives during a vacation period}\} \\
           {\times}&Pr\{\textrm{$A$ arrives during a vacation period}\} \\
           =&\frac{p_0U_0^{'}(1)}{p_0U_0^{'}(1)+(1-p_0)U_1^{'}(1)}{\times}(1-\lambda)
\label{xi0}
\end{split}
\end{equation}
\end{small}
and
\begin{equation}
Pr\{\xi=1\}=\frac{(1-p_0)U_1^{'}(1)}{p_0U_0^{'}(1)+(1-p_0)U_1^{'}(1)}{\times}(1-\lambda).
\label{xi1}
\end{equation}
On the other hand, the mean residual vacation period experienced by a packet given that it arrives during the vacation period $Y_i$ can be obtained via the analysis technique in \cite{pan2017TON} as follows:
\begin{equation}
E[R|\xi=i]=\frac{U_i^{''}(1)}{2{\lambda}U_i^{'}(1)}, i=0,1
\label{condi-R}
\end{equation}
Combining (\ref{xi0})-(\ref{condi-R}), the mean residual vacation period experienced by a packet is
\begin{equation}
\begin{split}
\overline{R}&=Pr\{\xi=0\}E[R|\xi=0]+Pr\{\xi=1\}E[R|\xi=1]  \\
            &=\frac{(1-\lambda)\left[p_0U_0^{''}(1)+(1-p_0)U_1^{''}(1)\right]}{2\lambda\left[p_0U_0^{'}(1)+(1-p_0)U_1^{'}(1)\right]}.
\label{meanR}
\end{split}
\end{equation}
Furthermore, based on (\ref{U1}) and (\ref{U0}), the mean residual vacation period will approach
\begin{equation}
\overline{R}=\frac{(1-\lambda)(1-\beta)}{\lambda\beta}=\frac{(1-\lambda)U_1^{'}(1)}{\lambda}=(1-\lambda)\overline{Y_1},
\label{meanR-2}
\end{equation}
if the number of nodes $n$ is sufficiently large.

Before a packet can get service, it may experience several complete vacation periods. As Fig.~\ref{fig10} illustrates, $A$ will see all the packets waiting outside the virtual gate if it arrives during a vacation period, and it may see that the packets both inside and outside the gate if it arrives during a busy period. Let $N_O$ be the number of packets waiting outside the gate upon the arrival of $A$. As Fig.~\ref{fig10} shows, if $A$ arrives during a busy period, it will experience $1+\left\lfloor\frac{N_O}{M}\right\rfloor$ complete vacation periods before it can be served; otherwise, it will undergo $\left\lfloor\frac{N_O}{M}\right\rfloor$ complete vacation periods. Using the technique in \cite{huangEPON}, we can obtain the average number of complete vacation periods that a packet experiences as follows:
\begin{equation}
\overline{E}=\lambda+\frac{\lambda\overline{W}}{M}-\frac{(1+\lambda)B^{''}(1)}{2MB^{'}(1)},
\label{vaca-num}
\end{equation}
where $B^{''}(1)$ is the difference between the second and the first moment of the busy period and can be derived via the relationship between $B$ and $Q$ in (\ref{Bt}) as follows:
\begin{equation}
\begin{split}
B^{''}(1)&=\sum_{k=1}^{M-1}k(k-1)q_k+\sum_{k=M}^{\infty}M(M-1)q_k  \\
         &=\frac{2\frac{G}{nr}\left[1-M\left(1-\frac{G}{nr}\left(\frac{G}{nr}\right)^{M-1}-\left(\frac{G}{nr}\right)^M\right)\right]}{\left(1-\frac{G}{nr}\right)^2}
\label{secMomentB}
\end{split}
\end{equation}
Also, since each vacation period completely experienced by a packet must start with a non-empty buffer, the expectation of such vacation periods is $\overline{Y_1}$. Thus, the average complete vacation periods experienced by a packet is given by
\begin{equation}
\overline{H}=\left[\lambda+\frac{\lambda\overline{W}}{M}-\frac{(1+\lambda)B^{''}(1)}{2MB^{'}(1)}\right]\overline{Y_1}.
\label{meanH}
\end{equation}
Combining (\ref{meanW}), (\ref{meanR}), and (\ref{meanH}), we can derive the formula of the mean waiting time as follows:
\newtheorem{myTheorem2}[myTheorem]{Theorem}
\begin{myTheorem2}
For a slotted Aloha with batch service, the mean waiting time of packets is given by
\begin{equation}
\begin{split}
\overline{W}&=\frac{\frac{(1-\lambda)\left[p_0U_0^{''}(1)+(1-p_0)U_1^{''}(1)\right]}{2\lambda\left[p_0U_0^{'}(1)+(1-p_0)U_1^{'}(1)\right]}+\overline{Y_1}\left[\lambda-\frac{(1+\lambda)B^{''}(1)}{2MB^{'}(1)}\right]}{1-\lambda-\frac{\lambda\overline{Y_1}}{M}}\\
            &\xrightarrow{large~n}\frac{\overline{Y_1}\left[1-\frac{(1+\lambda)B^{''}(1)}{2MB^{'}(1)}\right]}{1-\lambda-\frac{\lambda\overline{Y_1}}{M}}
\label{Theorem3}
\end{split}
\end{equation}
\end{myTheorem2}
\hfill$\blacksquare$

With the increase of $M$, the number of packets that can be transmitted in a busy period increases, thus the amortized competition overhead is small. Fig. 3 shows that the batch service can improve the network throughput remarkably. In the following, we study the delay performance of three cases where $M=1$, $M=2$, and $M=\infty$ using the result presented in Theorem 3 to demonstrate the batch service can also improve delay performance.

\subsection{Classical Slotted Aloha (M=1)}
When $M=1$, the slotted Aloha with batch service is actually the classical slotted Aloha. In this case, the busy period of each node is one slot, i.e.,
\begin{equation}
B^{'}(1)=1,
\label{meanB-M=1}
\end{equation}
and
\begin{equation}
B^{''}(1)=0.
\label{secondB-M=1}
\end{equation}

Also, Theorem 2 states that the attempt rate $G$ in this case satisfies the following equation
\begin{equation*}
\frac{\hat{\lambda}}{1-\hat{\lambda}}\left(\frac{1}{Ge^{-G}}-1\right)=1,
\end{equation*}
which changes to
\begin{equation}
Ge^{-G}=\hat{\lambda}
\label{G-M=1}
\end{equation}
after some reconfiguration. Thus, $G=-\mathbb{W}_{0}(-\hat{\lambda})$ or $G=-\mathbb{W}_{-1}(-\hat{\lambda})$. According to \cite{dai2012stability}, if the network is stable, the attempt rate $G$ eventually converges to $-\mathbb{W}_{0}(-\hat{\lambda})$. Using (\ref{mean-Y1}) and (\ref{G-M=1}), we thus have the mean vacation period $\overline{Y_1}$ for the classical slotted Aloha as follows:
\begin{equation}
\overline{Y_1}=\frac{1}{re^{-G}}-1=\frac{1}{re^{\mathbb{W}_{0}(-\hat{\lambda})}}-1.
\label{Y1-M=1}
\end{equation}
Substituting (\ref{meanB-M=1}), (\ref{secondB-M=1}), and (\ref{Y1-M=1}) into (\ref{Theorem3}), the mean waiting time of the classical slotted Aloha is given by
\begin{equation}
\overline{W}_{M=1}=\frac{1-re^{\mathbb{W}_{0}(-\hat{\lambda})}}{re^{\mathbb{W}_{0}(-\hat{\lambda})}-\lambda}
\label{W-M=1}
\end{equation}
To guarantee that the mean waiting time in (\ref{W-M=1}) is bounded when $\hat{\lambda}<e^{-1}$, the transmission probability $r$ should satisfy the following condition:
\begin{equation}
r>\frac{\lambda}{e^{\mathbb{W}_{0}(-\hat{\lambda})}}=\frac{-\mathbb{W}_{0}(-\hat{\lambda})}{n}.
\label{rdown-M=1}
\end{equation}
Also, the mean delay is bounded only if $\hat{\lambda}$ is equal to the network throughput $\hat{\lambda}_{out}$. According to (\ref{Theorem1}), the stable throughput region when $M=1$ is $r\in\left[\frac{-\mathbb{W}_{0}(-\hat{\lambda})}{n},\frac{-\mathbb{W}_{-1}(-\hat{\lambda})}{n}\right]$. Thus, the bounded delay region for the classical slotted Aloha is given by
\begin{equation}
r\in\left(\frac{-\mathbb{W}_{0}(-\hat{\lambda})}{n},\frac{-\mathbb{W}_{-1}(-\hat{\lambda})}{n}\right].
\label{BDR-M=1}
\end{equation}
Our analytical results (\ref{W-M=1}) and (\ref{BDR-M=1}) are consistent with that of the classical slotted Aloha in \cite{dai2012stability}, which implies that our model is a generalization of that in \cite{dai2012stability}.

The disadvantage of the classical slotted Aloha is that the bounded delay region shrinks remarkably with the increase of the number of nodes $n$ and the aggregate traffic $\hat{\lambda}$ \cite{abramson1973}, due to excessively large contention overhead. This can also be observed in (\ref{BDR-M=1}) and Fig. 11.

\begin{figure*}[!t]
\centering
\subfigure[$n=30$]{
\label{fig11-a}
\includegraphics[scale=0.35]{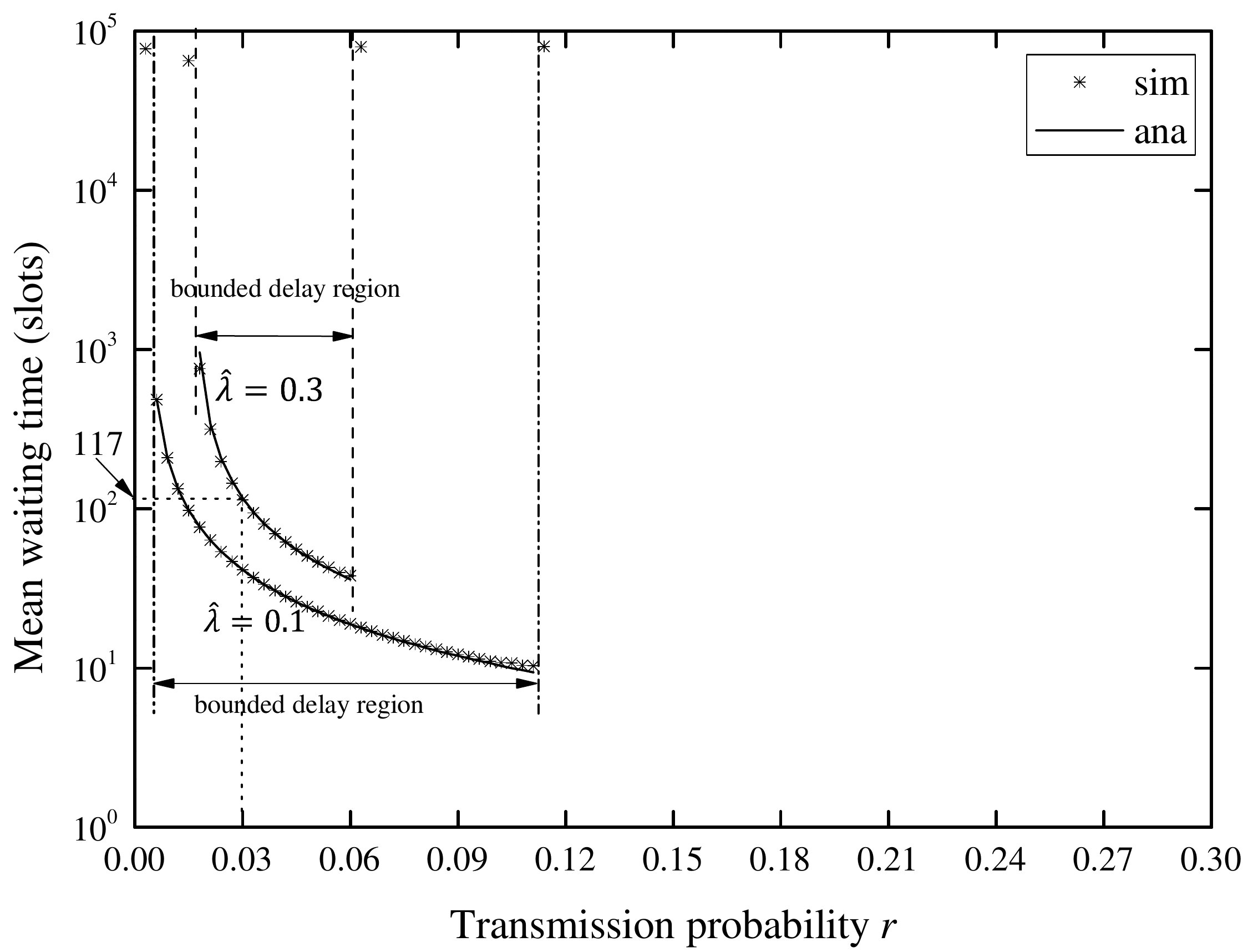}}
\subfigure[$n=50$]{
\label{fig11-b}
\includegraphics[scale=0.35]{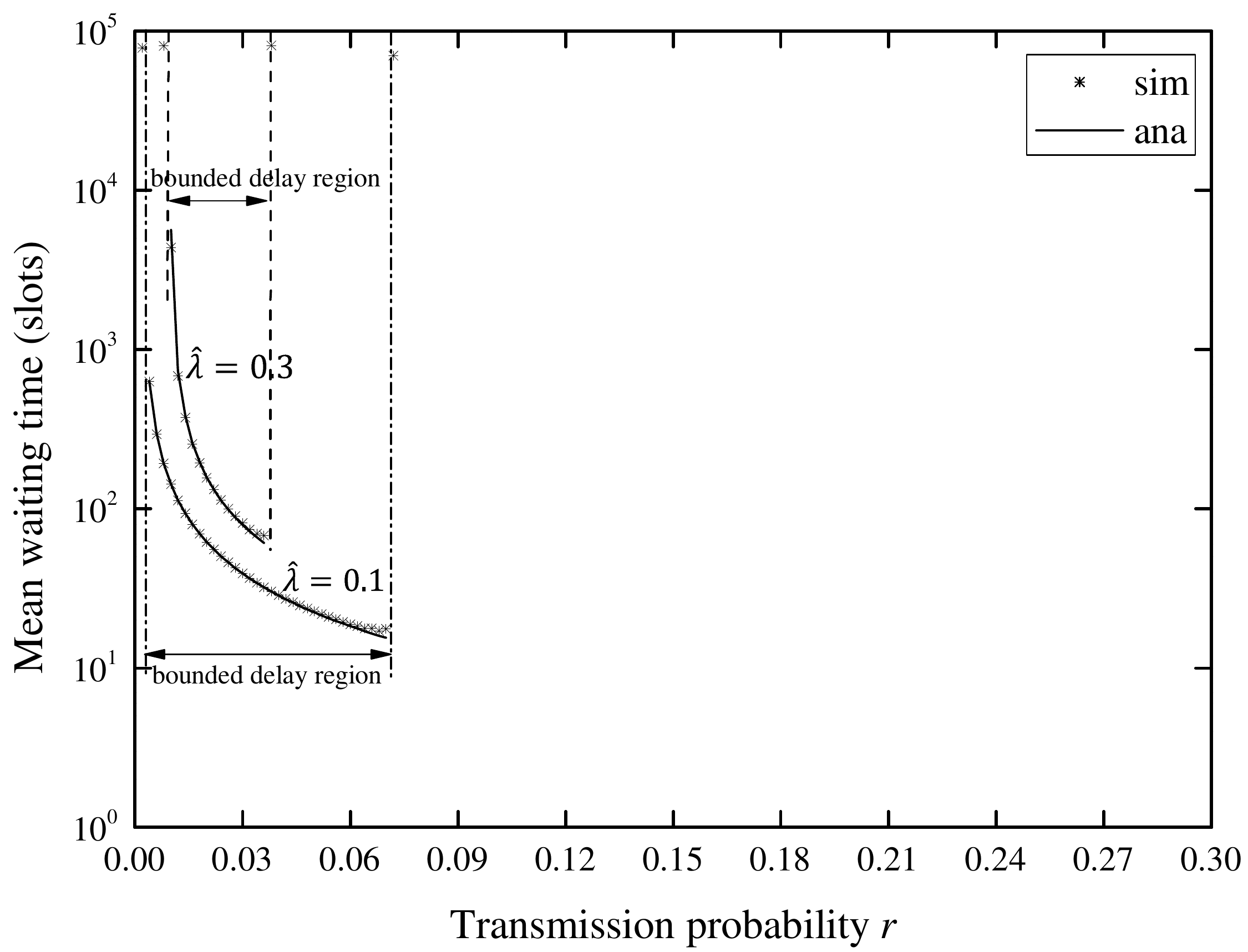}}
\caption{Mean waiting time and bounded delay region for classical slotted Aloha ($M=1$).}
\end{figure*}

\begin{figure*}[!t]
\centering
\subfigure[$n=30$]{
\label{fig12-a}
\includegraphics[scale=0.35]{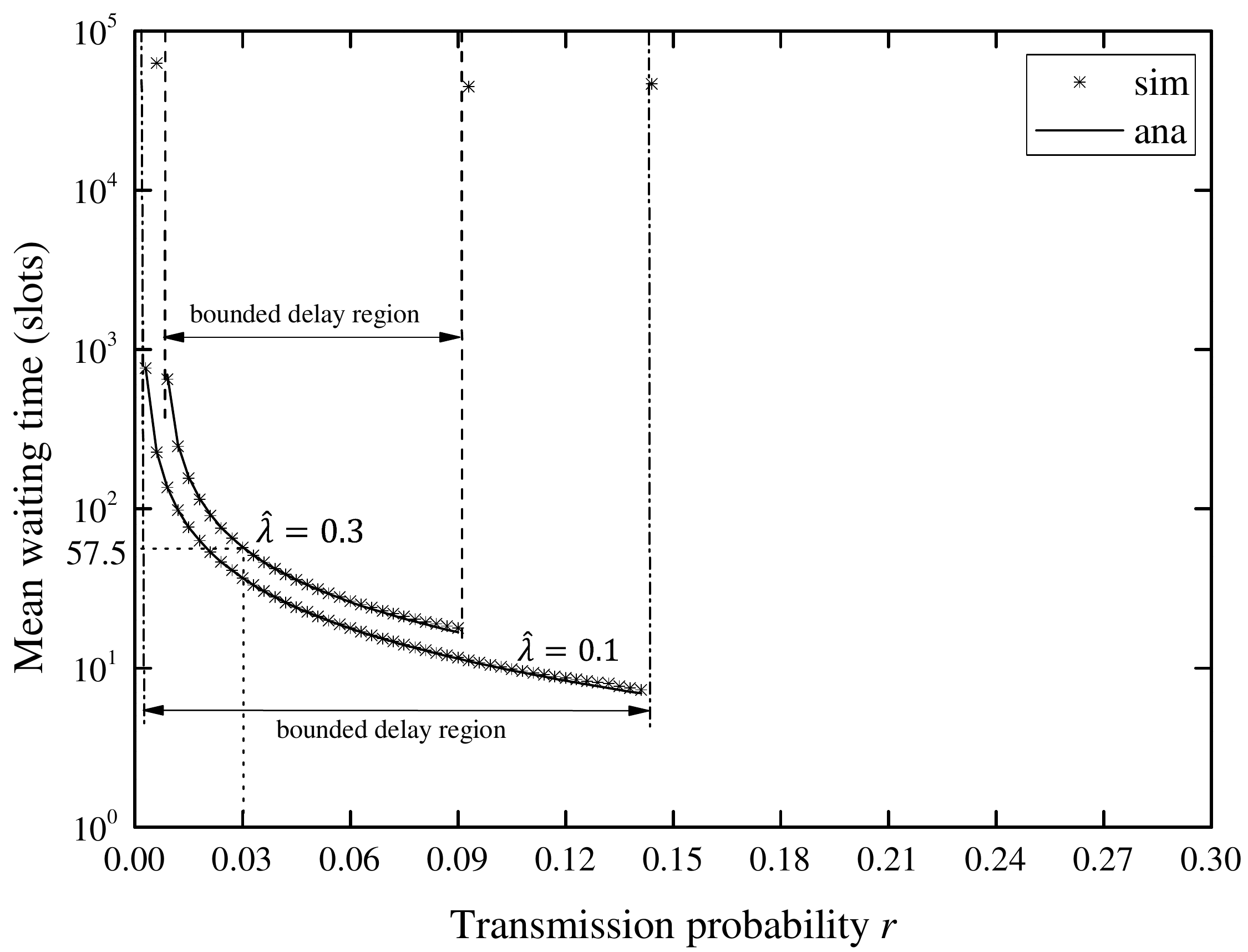}}
\subfigure[$n=50$]{
\label{fig12-b}
\includegraphics[scale=0.35]{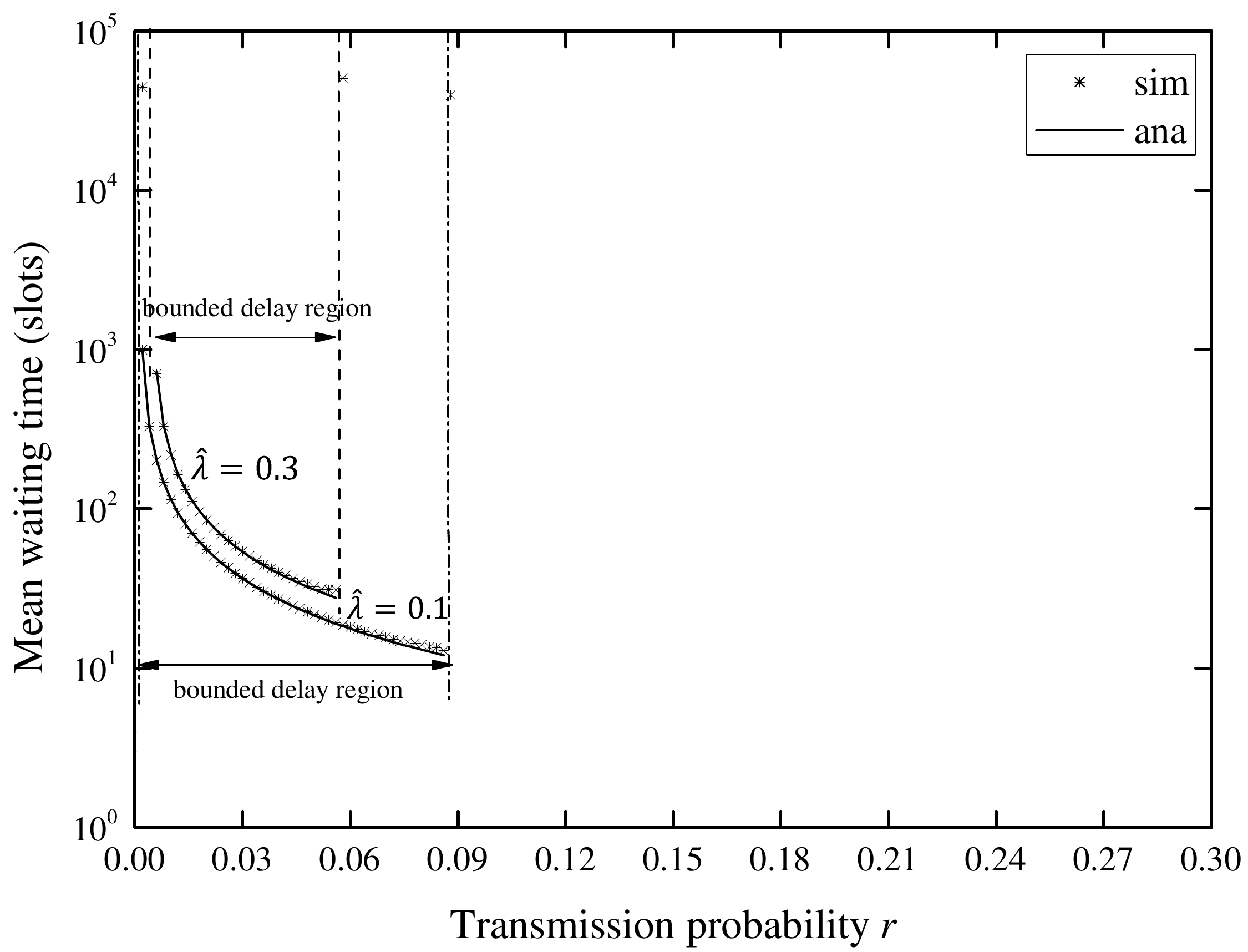}}
\caption{Mean waiting time and bounded delay region for batch-service slotted Aloha with $M=2$.}
\end{figure*}

\begin{figure*}[!t]
\centering
\subfigure[$n=30$]{
\label{fig13-a}
\includegraphics[scale=0.35]{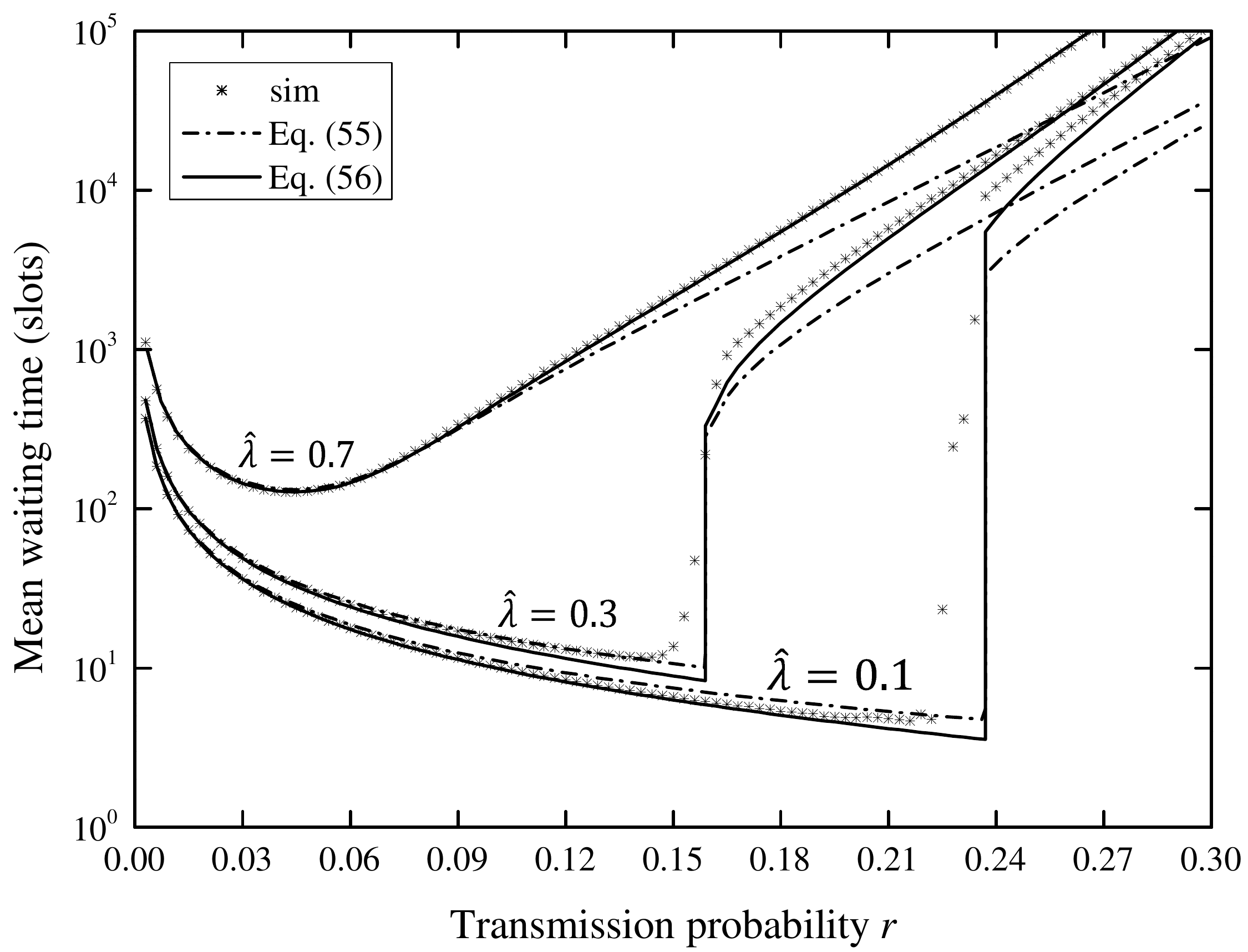}}
\subfigure[$n=50$]{
\label{fig13-b}
\includegraphics[scale=0.35]{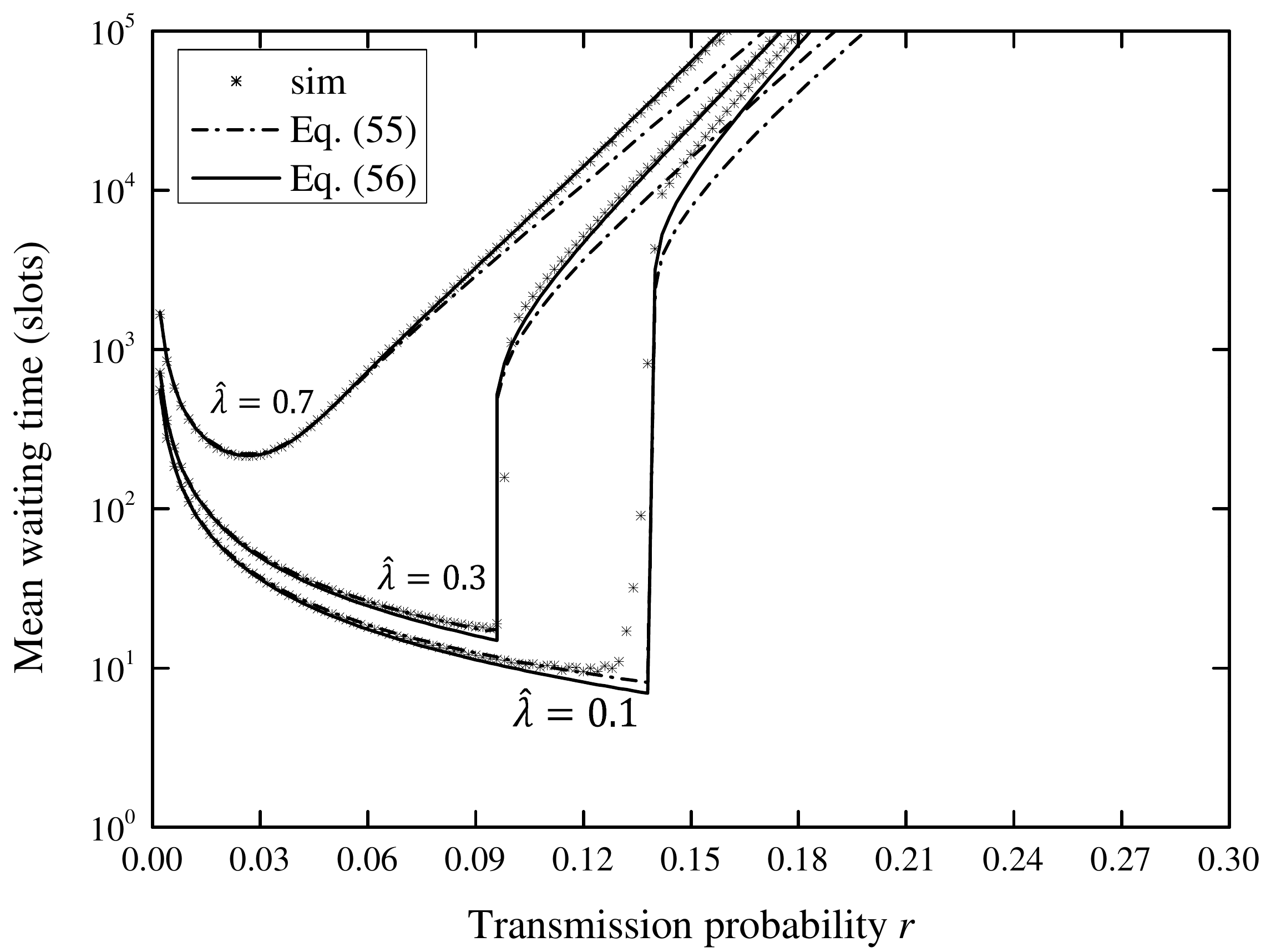}}
\caption{Mean waiting time and bounded delay region for batch-service slotted Aloha with $M=\infty$.}
\end{figure*}

\subsection{Batch-Service Slotted Aloha with $M=2$}
When $M$ increases from 1 to 2, the contention overhead of each packet decreases accordingly, i.e., each packet can be transmitted successfully after a smaller number of contentions. Intuitively, this may improve the delay performance of the slotted Aloha. To demonstrate this point, this subsection compares the batch service with $M=2$ to the classical slotted Aloha in terms of delay performance. We carry out the comparison numerically using the result presented in Theorem 3, since (\ref{Theorem2}) has no closed-form solution when $1<M<\infty$. In particular, we calculate the attempt rate $G$ by numerically solving (\ref{Theorem2}), and obtain the mean waiting time, denoted by $\overline{W}_{M=2}$, by substituting $G$ into (\ref{Theorem3}).

We plot $\overline{W}_{M=2}$ versus $r$ in Fig. 12, where $\hat{\lambda}=0.1$ packets/slot and 0.3 packets/slot, and $n=30$ and 50. As Fig. 12 shows, though the bounded delay region of the batch service with $M=2$ also shrinks with the increase of $\hat{\lambda}$ and $n$, it is larger than that of the classical slotted Aloha ($M=1$) if $\hat{\lambda}$ and $n$ are the same. For example, given $\hat{\lambda}=0.3$ and $n=30$, the bounded delay region of the classical slotted Aloha is $r\in(0.0163,0.0594]$ in Fig.~\ref{fig11-a}, while that of $M=2$ is now enlarged to $r\in(0.0073,0.0915]$ in Fig.~\ref{fig12-a}. Also, $\overline{W}_{M=2}$ is smaller than $\overline{W}_{M=1}$, given the values of $\hat{\lambda}$, $n$, and $r$. As an example, when $\hat{\lambda}=0.3$ packets/slot, $n=30$ and $r=0.03$, $\overline{W}_{M=1}=117$ slots while $\overline{W}_{M=2}=57.5$ slots.

\subsection{Batch-Service Slotted Aloha with $M=\infty$}
When $M=\infty$, once the node succeeds in the channel competition, it can send out all the packets waiting in the queue at the beginning of the slot when it succeeds. This implies that the queue length at the start of a busy period is equal to the length of this busy period. Therefore, we have
\begin{equation}
\overline{Q}=\overline{B}.
\label{Q-M=inf}
\end{equation}
Also, according to (\ref{Q}) and (\ref{mean-Y0}), there is
\begin{equation}
\begin{split}
\overline{Q}&=\overline{P}+\overline{U} \\
            &=\overline{P}+p_0\overline{U_0}+(1-p_0)\overline{U_1}  \\
            &=\overline{P}+p_0\lambda\overline{Y_0}+(1-p_0)\lambda\overline{Y_1} \\
            &=\overline{P}+p_0\lambda\left(\frac{1}{\lambda}+\overline{Y_1}\right)+(1-p_0)\lambda\overline{Y_1}  \\
            &=\overline{P}+p_0+\lambda\overline{Y_1}.
\label{meanQ-M=inf}
\end{split}
\end{equation}
When $M=\infty$, the queue length at the end of the busy period is equal to the number of arrivals during the busy period. Therefore, we have the following equations
\begin{equation}
\overline{P}=\overline{L}=\lambda\overline{B},
\label{meanP-M=inf}
\end{equation}
and
\begin{equation}
p_0=l_0=1-\lambda\overline{B},
\label{p0-M=inf}
\end{equation}
when $n$ is sufficiently large. Substituting (\ref{meanP-M=inf})-(\ref{p0-M=inf}) into (\ref{meanQ-M=inf}) and using (\ref{meanB}), we obtain
\begin{equation}
\overline{Y_1}=\frac{\overline{Q}-1}{\lambda}=\frac{\overline{B}-1}{\lambda}=\frac{1}{\lambda(1-\hat{\lambda})}\left(\frac{\hat{\lambda}}{Ge^{-G}}-1\right).
\label{Y1-M=inf}
\end{equation}
It follows from (\ref{Theorem3}) that the mean waiting time in this case is only determined by $\lambda$ and $\overline{Y_1}$. According to (\ref{Y1-M=inf}), we can formulate the mean waiting time under $M=\infty$ as a function of $G$ as follows:
\begin{equation}
\overline{W}_{M=\infty}=\frac{\overline{Y_1}}{1-\lambda}=\frac{1}{\lambda(1-\lambda)(1-\hat{\lambda})}\left(\frac{\hat{\lambda}}{Ge^{-G}}-1\right),
\label{W-M=inf}
\end{equation}
where the attempt rate $G$ is determined by (\ref{Theorem2}).

From (\ref{W-M=inf}), it is easy to find the superiority of the batch-service slotted Aloha with $M=\infty$ over that with finite $M$. Equation (\ref{W-M=inf}) clearly shows that the mean waiting time $\overline{W}_{M=\infty}$ is bounded for all the $r$s in the region (0,1) as long as the aggregate input traffic rate $\hat{\lambda}$ is smaller than 1. This indicates that, unlike the classical slotted Aloha and the batch service with $M=2$, the batch service with $M=\infty$ can be applied to the scenarios with large population sizes and high traffic loads. To confirm this point, we plot the mean waiting time $\overline{W}_{M=\infty}$ changing with $r$ in Fig. 13, where $\hat{\lambda}=0.1, 0.3$ and 0.7 packets/slot, and $n=30$ and 50. As Fig. 13 displays, regardless of $\hat{\lambda}$ and $n$, $\overline{W}_{M=\infty}$ is always bounded in the region $r\in(0,1)$.

Even so, the mean waiting time could be large when the transmission probability $r$ is relatively large. For example, given $\hat{\lambda}=0.3$ packets/slot and $n=30$, Fig.~\ref{fig13-a} shows that $\overline{W}_{M=\infty}$ is larger than $10^5$ slots if $r$ is set to 0.3. Thus, $r$ should be carefully tuned in practical applications to satisfy the delay requirement of users even when the batch service with $M=\infty$ is employed.

Fig. 13 also shows that the analytical results in (\ref{W-M=inf}) match the simulation results well when the transmission probability $r$ is small, but the divergence becomes prominent when $r$ is large. The error is mainly incurred by the assumption that the number of attempts in each slot obeys a Poisson distribution. Actually, when the number of nodes $n$ is finite, the number of attempts in each slot is binomially distributed, and thus the successful probability for an arbitrary node is
\begin{equation*}
np_{ne}r(1-p_{ne}r)^{n-1}=G\left(1-\frac{G}{n}\right)^{n-1},
\end{equation*}
which approaches $Ge^{-G}$ only when $n$ is very large and $r$ is very small. If we replace $Ge^{-G}$ with $G\left(1-\frac{G}{n}\right)^{n-1}$ in (\ref{W-M=inf}), we can obtain the following modified formula of the mean waiting time
\begin{equation}
\overline{W}_{M=\infty}=\frac{1}{\lambda(1-\lambda)(1-\hat{\lambda})}\left[\frac{\hat{\lambda}}{G\left(1-\frac{G}{n}\right)^{n-1}}-1\right].
\label{W-modi-M=inf}
\end{equation}
As Fig. 13 plots, the result in (\ref{W-modi-M=inf}) agrees with the simulation results in Fig. 13 very well when $r$ is relatively large.

\section{Conclusion}
We develop a generalized vacation model to study the performance of the slotted Aloha with batch service in this paper. Based on this model, we derive the delay performance of the slotted Aloha with batch service. Our analytical results show that the reduction of the amortized competition overhead induced by channel competition is the fundamental reason for the improvement of throughput and delay performances. Also, with the increase of batch size $M$, the system stable region in terms of the transmission probability $r$ is enlarged. Especially, when $M$ is infinity, the system stable region becomes the whole region of $r$, i.e., (0, 1) for any node population and any aggregate input traffic rate smaller than 1. This indicates the batch-service slotted Aloha with $M=\infty$ is quite robust with respect to $r$.

\bibliographystyle{IEEEtran}
\bibliography{IEEEabrv,IEEERef}

\begin{thebibliography}{10}
\providecommand{\url}[1]{#1}
\csname url@samestyle\endcsname
\providecommand{\newblock}{\relax}
\providecommand{\bibinfo}[2]{#2}
\providecommand{\BIBentrySTDinterwordspacing}{\spaceskip=0pt\relax}
\providecommand{\BIBentryALTinterwordstretchfactor}{4}
\providecommand{\BIBentryALTinterwordspacing}{\spaceskip=\fontdimen2\font plus
\BIBentryALTinterwordstretchfactor\fontdimen3\font minus
  \fontdimen4\font\relax}
\providecommand{\BIBforeignlanguage}[2]{{%
\expandafter\ifx\csname l@#1\endcsname\relax
\typeout{** WARNING: IEEEtran.bst: No hyphenation pattern has been}%
\typeout{** loaded for the language `#1'. Using the pattern for}%
\typeout{** the default language instead.}%
\else
\language=\csname l@#1\endcsname
\fi
#2}}
\providecommand{\BIBdecl}{\relax}
\BIBdecl

\bibitem{MTC2017}
D.~Vukobratovic, ``Massive machine-type communications and revival of
  {ALOHA},'' in \emph{INFOTEH-JAHORINA}, vol.~16, 2017.

\bibitem{SA2017ICC}
Q.~Song, X.~Lagrange, and L.~Nuaymi, ``An analytical model for {S-ALOHA}
  performance evaluation in {M2M} networks,'' in \emph{2017 IEEE International
  Conference on Communications (ICC)}.\hskip 1em plus 0.5em minus 0.4em\relax
  IEEE, 2017, pp. 1--7.

\bibitem{FASA2013TON}
H.~Wu, C.~Zhu, R.~J. La, X.~Liu, and Y.~Zhang, ``{FASA}: {A}ccelerated
  {S-ALOHA} using access history for event-driven {M2M} communications,''
  \emph{IEEE/ACM Transactions on Networking (ToN)}, vol.~21, no.~6, pp.
  1904--1917, 2013.

\bibitem{p-persistent2012}
D.~Wang, X.~Hu, F.~Xu, H.~Chen, and Y.~Wu, ``Performance analysis of {P-CSMA}
  for underwater acoustic sensor networks,'' in \emph{2012 Oceans-Yeosu}.\hskip
  1em plus 0.5em minus 0.4em\relax IEEE, 2012, pp. 1--6.

\bibitem{USN2011VTC}
Y.~Zhou, K.~Chen, J.~He, and H.~Guan, ``Enhanced slotted aloha protocols for
  underwater sensor networks with large propagation delay,'' in \emph{2011 IEEE
  73rd Vehicular Technology Conference (VTC Spring)}.\hskip 1em plus 0.5em
  minus 0.4em\relax IEEE, 2011, pp. 1--5.

\bibitem{802.15.6.2015}
S.~Rashwand, J.~Mi{\v{s}}i{\'c}, and V.~B. Mi{\v{s}}i{\'c}, ``Analysis of
  {CSMA/CA} mechanism of {IEEE} 802.15. 6 under non-saturation regime,''
  \emph{IEEE Transactions on parallel and Distributed Systems}, vol.~27, no.~5,
  pp. 1279--1288, 2015.

\bibitem{802.15.6standard}
``{IEEE} {S}tandard for {L}ocal and metropolitan area networks - part 15.6:
  {W}ireless {B}ody {A}rea {N}etworks,'' \emph{IEEE Std 802.15.6-2012}, pp.
  1--271, Feb 2012.

\bibitem{2005stability}
V.~Naware, G.~Mergen, and L.~Tong, ``Stability and delay of finite-user slotted
  {ALOHA} with multipacket reception,'' \emph{IEEE Transactions on Information
  theory}, vol.~51, no.~7, pp. 2636--2656, 2005.

\bibitem{kleinrock1973}
L.~Kleinrock and S.~S. Lam, ``Packet-switching in a slotted satellite
  channel,'' in \emph{Proceedings of the June 4-8, 1973, national computer
  conference and exposition}.\hskip 1em plus 0.5em minus 0.4em\relax ACM, 1973,
  pp. 703--710.

\bibitem{SA2017distributed}
W.~T. Toor, J.-B. Seo, and H.~Jin, ``Distributed transmission control in
  multichannel {S}-{ALOHA} for ad-hoc {N}etworks,'' \emph{IEEE Communications
  Letters}, vol.~21, no.~9, pp. 2093--2096, 2017.

\bibitem{Bayesian2013}
J.-B. Seo, H.~Jin, and V.~C. Leung, ``A terminal-assisted {B}ayesian
  broadcasting algorithm for {S-ALOHA} systems with finite population of
  multi-buffered terminals,'' \emph{IEEE Communications Letters}, vol.~17,
  no.~11, pp. 2064--2067, 2013.

\bibitem{2001pseudoBayesian}
J.-F. Frigon and V.~C. Leung, ``A pseudo-{B}ayesian {ALOHA} algorithm with
  mixed priorities,'' \emph{Wireless Networks}, vol.~7, no.~1, pp. 55--63,
  2001.

\bibitem{SA2017TON}
L.~Barletta, F.~Borgonovo, and I.~Filippini, ``The {T}hroughput and {A}ccess
  {D}elay of {S}lotted-{A}loha with {E}xponential {B}ackoff,'' \emph{IEEE/ACM
  Transactions on Networking}, vol.~26, no.~1, pp. 451--464, 2017.

\bibitem{2005BackoffTON}
B.-J. Kwak, N.-O. Song, and L.~E. Miller, ``Performance analysis of exponential
  backoff,'' \emph{IEEE/ACM Transactions on Networking (TON)}, vol.~13, no.~2,
  pp. 343--355, 2005.

\bibitem{M2M2018Dai}
W.~Zhan and L.~Dai, ``Massive random access of machine-to-machine
  communications in {LTE} networks: Modeling and throughput optimization,''
  \emph{IEEE Transactions on Wireless Communications}, vol.~17, no.~4, pp.
  2771--2785, 2018.

\bibitem{jiang2018mIoT}
N.~Jiang, Y.~Deng, A.~Nallanathan, X.~Kang, and T.~Q. Quek, ``Analyzing random
  access collisions in massive {I}o{T} networks,'' \emph{IEEE Transactions on
  Wireless Communications}, vol.~17, no.~10, pp. 6853--6870, 2018.

\bibitem{T-Lohi}
A.~A. Syed, W.~Ye, and J.~Heidemann, ``Comparison and evaluation of the
  {T}-{L}ohi {MAC} for underwater acoustic sensor networks,'' \emph{IEEE
  Journal on Selected Areas in Communications}, vol.~26, no.~9, pp. 1731--1743,
  2008.

\bibitem{Dai2019random}
Y.~Gao and L.~Dai, ``Random access: Packet-based or connection-based?''
  \emph{IEEE Transactions on Wireless Communications}, vol.~18, no.~5, pp.
  2664--2678, 2019.

\bibitem{abramson1973}
N.~Abramson, ``Packet switching with satellites.'' HAWAII UNIV HONOLULU, Tech.
  Rep., 1973.

\bibitem{1999stability}
W.~Luo and A.~Ephremides, ``Stability of {$N$} interacting queues in
  random-access systems,'' \emph{IEEE Transactions on Information Theory},
  vol.~45, no.~5, pp. 1579--1587, 1999.

\bibitem{1988stability}
R.~R. Rao and A.~Ephremides, ``On the stability of interacting queues in a
  multiple-access system,'' \emph{IEEE Transactions on Information Theory},
  vol.~34, no.~5, pp. 918--930, 1988.

\bibitem{1979ergodicity}
B.~S. Tsybakov and V.~A. Mikhailov, ``Ergodicity of a slotted {ALOHA} system,''
  \emph{Problemy peredachi informatsii}, vol.~15, no.~4, pp. 73--87, 1979.

\bibitem{dai2012stability}
L.~Dai, ``Stability and delay analysis of buffered {A}loha networks,''
  \emph{IEEE Transactions on Wireless Communications}, vol.~11, no.~8, pp.
  2707--2719, 2012.

\bibitem{Dai2013CSMA}
------, ``Toward a coherent theory of {CSMA} and {A}loha,'' \emph{IEEE
  Transactions on Wireless Communications}, vol.~12, no.~7, pp. 3428--3444,
  2013.

\bibitem{sun2017fairness}
X.~Sun and L.~Dai, ``Fairness-constrained maximum sum rate of multi-rate csma
  networks,'' \emph{IEEE Transactions on Wireless Communications}, vol.~16,
  no.~3, pp. 1741--1754, 2017.

\bibitem{wong2011CSMA}
P.~K. Wong, D.~Yin, and T.~T. Lee, ``Analysis of non-persistent {CSMA}
  protocols with exponential backoff scheduling,'' \emph{IEEE Transactions on
  Communications}, vol.~59, no.~8, pp. 2206--2214, 2011.

\bibitem{lambertw}
R.~M. Corless, G.~H. Gonnet, D.~E. Hare, D.~J. Jeffrey, and D.~E. Knuth, ``On
  the {L}ambert{W} function,'' \emph{Advances in Computational mathematics},
  vol.~5, no.~1, pp. 329--359, 1996.

\bibitem{pan2017TON}
X.~Pan, T.~Ye, T.~T. Lee, and W.~Hu, ``Power efficiency and delay tradeoff of
  10{GB}ase-{T} energy efficient {E}thernet protocol,'' \emph{IEEE/ACM
  Transactions on Networking}, vol.~25, no.~5, pp. 2773--2787, 2017.

\bibitem{huangEPON}
H.~Huang, T.~Ye, and T.~T. Lee, ``Optimum {T}ransmission {W}indow for {EPON}s
  with {L}imited {S}ervice,'' \emph{IEEE Access}, vol.~7, pp. 57\,956--57\,971,
  2019.

\end{thebibliography}
\end{document}